\documentclass[journal]{IEEEtran}
\pdfoutput=1

\usepackage{cite}
\usepackage[]{footmisc}

\usepackage[pdftex]{graphicx}

\usepackage{booktabs,colortbl}
\newcommand{\ra}[1]{\renewcommand{\arraystretch}{#1}}

\usepackage{subfigure}
\graphicspath{{./img/}}
\DeclareGraphicsExtensions{.pdf}
\ifCLASSINFOpdf
\else
 
\fi
\usepackage{amsmath, amssymb}
\usepackage{color, soul}
\usepackage{algorithm}
\PassOptionsToPackage{noend}{algpseudocode}
\usepackage{algpseudocode}

\usepackage{mdwtab}

\ifodd 0
\newcommand{\rev}[1]{{\color{blue}#1}} 

\newcommand{\com}[1]{\textbf{\color{red}([QZ]: #1)}}
\newcommand{\comMC}[1]{\textbf{\color{red}([MC]: #1)}}
\newcommand{\comCW}[1]{\textbf{\color{red}([CW]: #1)}}
\newcommand{\comMW}[1]{\textbf{\color{red}([MW]: #1)}}
\else
\newcommand{\rev}[1]{#1}

\newcommand{\com}[1]{}
\newcommand{\comMC}[1]{}
\newcommand{\comCW}[1]{}
\newcommand{\comMW}[1]{}
\fi

\begin{document}

\title{Adaptive Multi-Trace Carving
for\\ Robust Frequency Tracking in Forensic Applications}

\author{Qiang~Zhu,~\IEEEmembership{Student Member,~IEEE,}  
		Mingliang~Chen,~\IEEEmembership{Student Member,~IEEE,}
		Chau-Wai~Wong,~\IEEEmembership{Member,~IEEE,} 
        and~Min~Wu,~\IEEEmembership{Fellow,~IEEE}
\thanks{Q. Zhu was with the Department of Electical and Computer Engineering, University of Maryland, Collge Park, MD, 20742 USA, where the work was carried out, and is now with Facebook Inc. E-mail: zhuqiang@terpmail.umd.edu.}
\thanks{M. Chen, and M. Wu are with the Department
of Electrical and Computer Engineering, University of Maryland, Collge Park,
MD, 20742 USA. E-mail: \{mchen126, minwu\}@umd.edu.}
\thanks{C.-W. Wong was with the Department of Electrical and Computer Engineering, University of Maryland, College Park, MD, 20742 when this work was started, and is now with the Department of Electrical and Computer Engineering, North Carolina State University, Raleigh, NC, 27695 USA. E-mail: chauwai.wong@ncsu.edu.}
\thanks{A preliminary version reporting early-stage results of this work was presented in the 2018 Asilomar Conference on Signals, Systems, and Computers~\cite{zhu2018AMTCAsilomar}.}}

\maketitle

\begin{abstract}
In the field of information forensics, many emerging problems involve a critical step that estimates and tracks weak frequency components in noisy signals. It is often challenging for the prior art of frequency tracking to i)~achieve a high accuracy under noisy conditions, ii)~detect and track multiple frequency components efficiently, or iii)~strike a good trade-off of the processing delay versus the resilience and the accuracy of tracking. To address these issues, we propose Adaptive Multi-Trace Carving (AMTC), a unified approach for detecting and tracking one or more subtle frequency components under very low signal-to-noise ratio (SNR) conditions and in near real time. AMTC takes as input a time-frequency representation of the system's preprocessing results (such as the spectrogram), and identifies frequency components through iterative dynamic programming and adaptive trace compensation. The proposed algorithm considers relatively high energy traces sustaining over a certain duration as an indicator of the presence of frequency/oscillation components of interest and track their time-varying trend. Extensive experiments using both synthetic data and real-world forensic data of power signatures and physiological monitoring reveal that the proposed method outperforms representative prior art under low SNR conditions, and can be implemented in near real-time settings. The proposed AMTC algorithm can empower the development of new information forensic technologies that harness very small signals.
\end{abstract}


\IEEEpeerreviewmaketitle

\section{Introduction}
\IEEEPARstart{T}{he} recent two decades have seen a rapid growth of digital information forensic research~\cite{stamm2013InfoForensicReview}, with applications from tampering detection, to spatial-temporal verification, to more recently, physiological forensic analysis. Many of these emerging information forensic problems involving small and noisy signals include a critical step of estimating and tracking the instantaneous frequency or oscillation rate. Examples include the imperceptible environmental frequency traces such as power signatures in the form of Electric Network Frequency (ENF) signals~\cite{garg2013seeing,su2014ENFfromVidRollingShutter,su2014enfvideosync}, and the pulse frequency traces in the form of remote photoplethysmogram from facial videos~\cite{verkruysse2008FirstrPPG,liu2018DeepRppgFaceAntiSpoofing,li2016rPPGFaceAntiSpoofing}.

As the extraction of frequency traces often plays a key role in the aforementioned forensic applications, one needs to carefully answer the following questions before deploying a frequency estimator:
\begin{enumerate}
    \item Can the frequency components be detected from the digital recording?
    \item If a frequency component is detected, can the frequency be accurately estimated, especially in low signal-to-noise ratio (SNR) conditions?
\end{enumerate}
Answering the above problems can be challenging due to the relatively low signal strength of the components-of-interest compared with other audio or visual contents in the recording. To successfully estimate the frequency of interest within the noisy signal, an algorithm must be robust under strong noise and have the capability to exclude strong interferences.
\par In this paper, we take as input the time-frequency representation of the system's preprocessing results, such as a spectrogram, to perform frequency estimation and tracking. We propose a detection and tracking method for multiple frequency traces based on iterative dynamic programming and adaptive trace compensation. Inspired by the seam carving algorithm for content-aware image resizing~\cite{avidan2007seam}, we relate the problem of finding a smooth frequency trace to that of finding the trace of maximum energy in a spectrogram. Considering the inherent continuity in many forensic problems, we incorporate an additional temporal regularization term that favors close frequency estimates in consecutive time bins. Such a problem can be efficiently solved using a dynamic programming framework. 

\par
\begin{figure}
\centering
\subfigure[]{\includegraphics[width=1.6in]{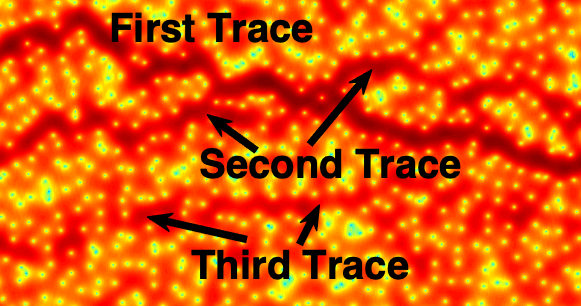}
\label{subfig_ExpSynthOri}}
\subfigure[]{\includegraphics[width=1.6in]{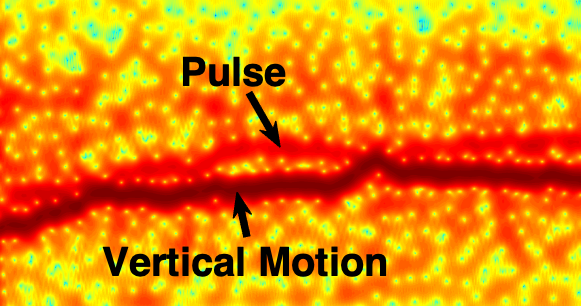}
\label{subfig_ExpSynthRes}}
\subfigure[]{\includegraphics[width=1.6in]{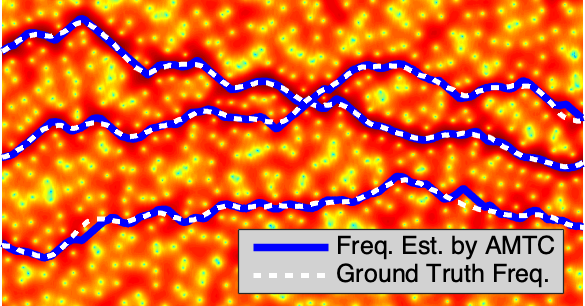}
\label{subfig_ExpHROri}}
\subfigure[]{\includegraphics[width=1.6in]{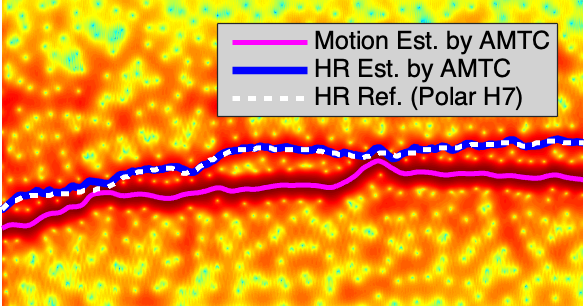}
\label{subfig_ExpHRRes}}
\caption{(a) Spectrogram of a synthetic $-10$~dB signal with three  frequency components and (c) the same image overlaid with ground-truth frequency components (white dashed line) and the frequency estimates using AMTC (blue line). (b) Spectrogram of a remote-photoplethysmogram signal with a weak pulse trace masked by a strong trace induced by the motion of a subject exercising on an elliptical machine~\cite{zhu2017ICIP} and (d) the spectrogram overlaid with pulse rate estimate (blue line) after compensating the first trace (magenta line) using AMTC. The estimation result is compared with the heart rate (white dashed line) simultaneously measured by an electrocardiogram-based sensor.}
\label{fig_introdemo}
\end{figure}

 Considering the presence of multiple traces within the frequency range of interest is possible, we propose an iterative frequency tracking method named \textit{Adaptive Multi-Trace Carving} (AMTC) to track all candidate traces. We first apply a proposed single-frequency tracking method to obtain the dominating frequency and then compensate the energy of the previous trace at the end of each iteration to facilitate the estimation of a next trace. Fig.~\ref{fig_introdemo} shows two tracking examples using AMTC on highly-corrupted signals, where the estimation results of AMTC are almost identical to the references. An efficient near real-time algorithm is also proposed by assuming the Markovian property of traces and introducing a bidirectional time window. We call it the online-AMTC. Although we mainly consider the spectrogram in this paper, our proposed techniques can be applied to other time-frequency representations of the signal for which the temporal tracking of signal traces is needed~\cite{zhang2019smars}.

The main contributions of this work are as follows:
\begin{enumerate}
\item For the task of the frequency-based media forensics, we have proposed a robust approach for frequency detection and tracking which can accurately and efficiently track multiple frequency traces in very low SNR conditions (usually $\leq -10$~dB). This method does not require the prior knowledge of the signal's specific statistical characteristics.
\item We adapt our proposed baseline algorithm of offline-AMTC into an efficient near--real-time algorithm. We reduce the computational complexity with a queuing data structure and maintain the performance level comparable with the offline version.
\item We conduct extensive experiments and analysis using challenging synthetic data and real-world forensic data. Several estimation methods initially proposed for other applications (such as the pitch estimation) are implemented, re-trained (the factorial hidden Markov model based method~\cite{wohlmayr2011probabilistic}), and compared. The results in Section~\ref{section_exp_subsection_simu} will show that the proposed offline-AMTC outperforms the Particle Filter method~\cite{shi2003spectrogram} and the YAAPT method~\cite{kasi2002yet} (which integrated normalized cross correlations, spectrogram peaks, and dynamic programming) in terms of the accuracy in a single-trace tracking scenario, and outperforms the factorial hidden Markov model based method~\cite{wohlmayr2011probabilistic} in terms of the accuracy and efficiency in a multi-trace tracking scenario.
\item We present a novel detection method based on the AMTC framework to accurately test the presence of trace and discuss other considerations when using the approach, such as the estimation of the number of frequency components and the accommodation of human-in-the-loop involvement.
\end{enumerate}

\par The rest of the paper is organized as follows. In Section~\ref{section_RelatedWork}, the background information and the related work about the frequency tracking problems are discussed. In Section~\ref{section_Problem}, we formulate the problem of single-trace tracking and solve it using dynamic programming. In Section~\ref{subsection_offlineAMTC}, we propose the offline multi-trace tracking method or the offline-AMTC, based on an iterative and greedy search strategy. In Section~\ref{subsection_onlineAMTC}, we present the online-AMTC. In Section~\ref{section_exp}, we show the experimental results comparing the performance of AMTC with several representative prior methods using both synthetic and real-world data. In Section~\ref{sec::impact_factors}, we evaluate the impact of various factors on the performance. In Section~\ref{section_discussion}, we discuss several practical issues as well as the limitations of the AMTC algorithms. Section~\ref{section_conclude} concludes the paper.

\section{Background and Related Works}\label{section_RelatedWork}

\subsection{Micro-Signal Extraction}
A number of information forensic challenges often boil down to the frequency extraction problem, where the signals-of-interest have smaller amplitude or size, typically by about one order of magnitude, than the dominating or hosting signals~\cite{CWWong2017MicroSig}. Among the forensic applications involving micro-signals, the Electric Network Frequency (ENF) signals and the video-based remote-photoplethysmography (rPPG) are two emerging examples. Here, we briefly discuss these two applications that inspired this work and will serve as the source of real-world data examples to demonstrate our proposed algorithms.

ENF signal can be captured by audio recordings made near mains-powered appliances due to electromagnetic interference, acoustic hum, and/or mechanical vibrations. ENF signal can also be captured by photodiodes and cameras due to ENF-induced flicking of mains-powered light sources. As the ENF variation at each time instant and location differs from each other, the recording time and place can be validated by matching the ENF extracted from the recording with the reference obtained from the power mains.

The rPPG technique is an emerging approach to tackle face sproofing and forgeries~\cite{li2016rPPGFaceAntiSpoofing,liu2018DeepRppgFaceAntiSpoofing}. It has been shown possible in~\cite{verkruysse2008FirstrPPG} to extract a person's instantaneous pulse rate (PR) from his/her facial video by examining the subtle pulse-induced color change of the facial skin pixels, including when the video contains significant movement and environmental illumination changes~\cite{zhu2017ICIP,wang2015exploiting}. rPPG based physiological forensics are showing promise in sports, fitness, as well as public health~\cite{wu2020physioForensics}.

In these forensic applications, the presence of multiple frequency traces within a certain frequency range is possible. For the ENF scenario, strong acoustic interference from other sources may very well dominate the weak ENF traces. For the rPPG scenario, the frequency trace resulting from a person's movement (such as running on a treadmill) may coexist with the pulse frequency trace, as seen in Fig. 1(b). Using a multi-trace search strategy increases the chance of finding the appropriate trace-of-interest. 
In addition, being able to run a tracking algorithm in real time may also be important for such applications as physiological sensing.

\subsection{Prior Art on Frequency Tracking}
\par Traditional frequency estimation algorithms are often applied individually to each temporal segment, assuming segment-wise signal stationarity. Subspace methods such as multiple signal classification (MUSIC) \cite{schmidt1986multiple} and estimation of signal parameters via rotational invariance technique (ESPIRIT) \cite{roy1989esprit} build pseudo power spectra using parametric models of pure sinusoids. These frame-wise estimation algorithms do not explicitly exploit the temporal correlation of neighboring segments and become less accurate as the SNR drops and frequently generate outliers. 
\par The problem of tracking a single frequency component has been extensively studied. In~\cite{shi2003spectrogram}, a sequential Monte Carlo method was proposed, and importance sampling was used to approximate the posterior distribution of each frequency state. However, without a backward smoothing procedure, the output tracking results tend to be inaccurate when substantial interference exists, and the resampling stage makes the algorithm time-consuming. In~\cite{streit1990frequency}, a prior knowledge of trace dynamic was utilized, and the problem was formulated as a hidden Markov model (HMM) problem. The maximum a posteriori probability estimate was efficiently calculated by running a Viterbi solver. However, HMM requires both the modeling and calibration of a key building block, the emission probability. Such a pre-calibration requirement often makes this method hard to be deployed in real-world forensic tasks, especially when the training data is unavailable. The recently developed Yet Another Algorithm for Pitch Tracking (YAAPT)~\cite{zahorian2008spectral} focused on single pitch estimation of speech signal based on both spectrogram and correlogram. The authors proposed using dynamic programming to estimate the fundamental frequency trace from a set of candidate peaks of proposed harmonic spectral features. A similar tracking method can be found in~\cite{ojowu2012enf_tracking}. Such local-peak based methods guarantee excellent performance in high SNR cases, but often generate biased estimates under low SNR, as the probability that a local peak represents the actual signal frequency drops significantly. 
\par The problem of tracking multiple frequency components from the spectrogram image has also been investigated. Image processing techniques such as morphological operators \cite{abel1992image} and active contour~\cite{lampert2010active} methods have been applied to this area, but these methods may be difficult to be adapted to real-time tracking algorithms. Wohlmayr \textit{et al.} \cite{wohlmayr2011probabilistic} modeled the probability of pitch using Gaussian mixture models (GMMs), and then used the junction tree algorithm to decode a speaker-dependent factorial HMM (fHMM). A similar approach can be found in \cite{liu2017speaker}, where the emission probability was modeled by a deep neural network (DNN). Although both methods provide excellent performance in terms of accuracy for speech analysis, it is not always possible to meet the general needs in the real world for the following two reasons. First, the training phase requires a large amount of real-world data, which is often unavailable for most tasks beyond speech applications. Second, it is relatively time-consuming to compute the frame-wise joint emission probability and to decode the fHMM with the junction tree algorithm. More recent studies~\cite{review22019LP,review22018LP} proposed to use linear programming to find the best connection path of the frequency peaks on the spectrogram. These two methods first obtain all frequency peaks in the spectrogram as candidates and then find the best path from the candidates via linear programming. For low SNR scenarios, such approaches may find a large number of frequency peaks as the candidates, leading to huge memory and computational cost that is not scalable.

\begin{figure*}
\centering
\includegraphics[width=7in]{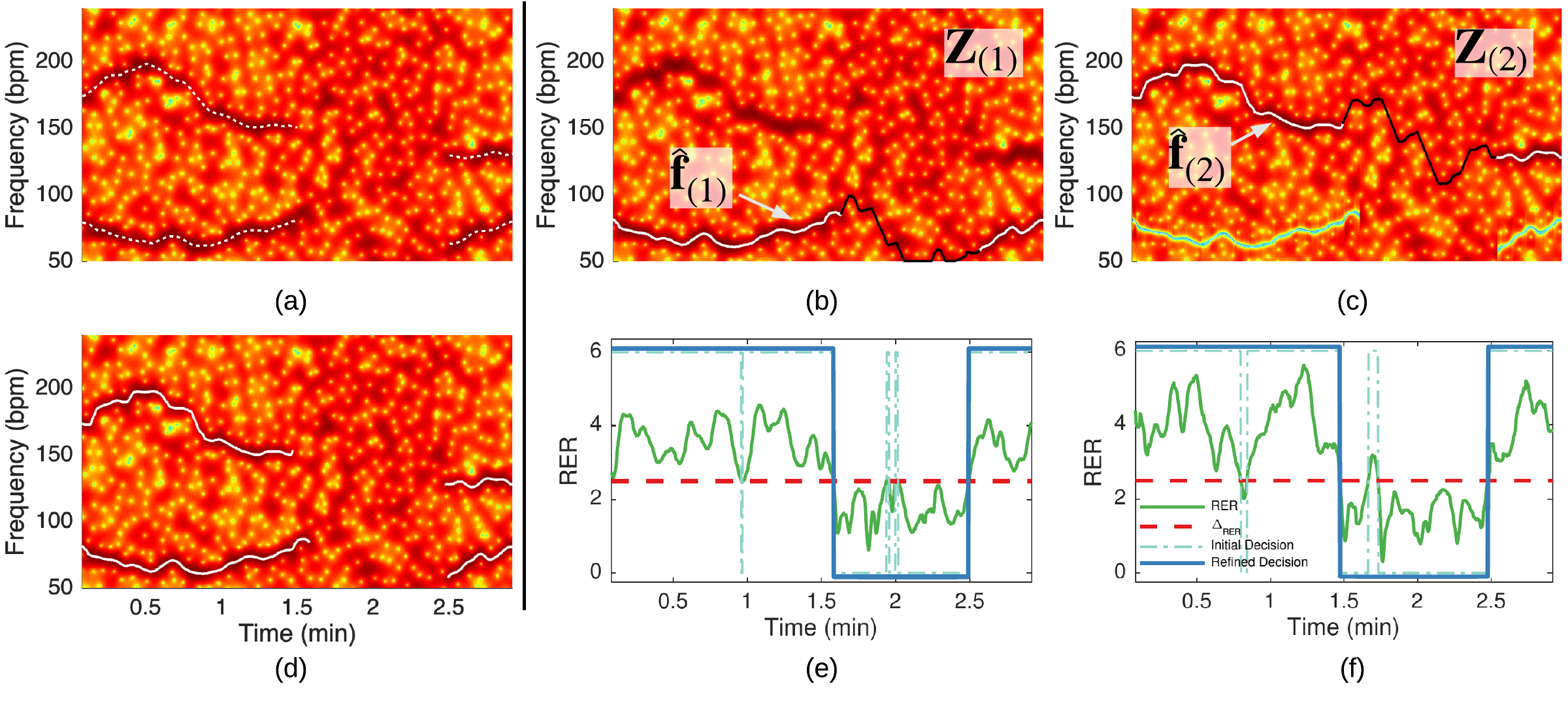}
\caption{Illustrations for an offline-AMTC estimation process: (a) spectrogram of a synthetic $-8$~dB signal with two frequency components. The unvoiced segment is from $1.5$ to $2.5$ min (white dots: ground truth); (b) first and (c) second trace estimates in voiced decision regions (white line) and unvoiced decision regions (black line) by AMTC; (e)--(f) test statistic RER and the corresponding voiced decision; (d) final trace estimate.}
\label{fig_OfflineAMTCExample01}
\end{figure*}

\section{Tracking a Single Frequency Trace}\label{section_Problem}
In this section, we present a trace tracking method which provides a practical and robust solution for tracking a single frequency trace. We formulate the problem by taking into account the energy as well as the smoothness of the trace. We adopt a dynamic programming algorithm to efficiently search for a candidate trace. A trace presence test is applied at the frame level to finalize the estimated trace.
\subsection{Problem Formulation}
We first formulate a frequency tracking problem for the scenarios that only a single trace exists in a frequency range of interest. Let $\mathbf{Z}\in \mathbb{R}_+^{M\times N}$ be the magnitude of a signal spectrogram image, which has $N$ discretized bins along the time axis and $M$ bins along the frequency axis. We define a \textit{frequency trace} as
\begin{equation}
\mathbf{f} = \{(f(n), n)\}_{n = 1}^N,
\end{equation}
where $f$: $[1,N]\rightarrow [1,M]$ is a function. Given the spectrogram $\mathbf{Z}$ and a candidate trace $\mathbf{f}$, we define an energy function for the trace as $E(\mathbf{f})= \sum^N_{n=1}\mathbf{Z}(f(n), n)$. A reasonable estimate of the frequency trace for the given signal is the trace $\hat{\mathbf{f}}$ that maximizes the energy function shown as follows

\begin{equation}
\begin{aligned}
& \hat{\mathbf{f}}=\underset{\mathbf{f}}{\text{argmax}}
& & E(\mathbf{f}).\\
\end{aligned}\label{P_general}
\end{equation}
Problem~(\ref{P_general}) is equivalent to the peak-finding method \cite{rife1974single,garg2013seeing} where $\hat{f}(n)= \underset{f(n)}{\text{argmax}}\,\mathbf{Z}(f(n),n)$, $\forall n \in [1,N]$. It also shares a similar spirit as the weighted average approach \cite{garg2013seeing}. 

\par To take into consideration the smoothness assumption of the trace along time, we add a regularization term that penalizes jumps in the frequency value. We model the change of the frequency value between two consecutive bins at $n-1$ and $n$ as a one step discrete-time Markov chain, characterized by the prior distribution function $P_m$ and the transition probability matrix $\mathbf{P}\in \mathbb{R}^{M\times M}$, where $P_m= P(f(1)= m)$ and $P_{m'm} = P(f(n)=m|f(n-1)=m')$, $\forall m,m' = 1, ..., M$, and $\forall n= 2,...,N$. Note that we assume $P_m$ to be uniformly distributed throughout this paper to treat the initial presence of each frequency state equally, even though it is possible to use other choices based on the available prior knowledge. The regularized single-trace frequency tracking problem is formulated as follows

\begin{equation}
\begin{aligned}
& \hat{\mathbf{f}}=\underset{\mathbf{f}}{\text{argmax}}
& & E(\mathbf{f})+\lambda P(\mathbf{f}), \\
\end{aligned}\label{P_regularized}
\end{equation}
where $P(\mathbf{f}) \triangleq \log P(f(1))+ \sum_{n=2}^N\log P(f(n)|f(n-1))$, and $\lambda>0$ is a regularization parameter that controls the smoothness of the resulting trace. 

\subsection{Efficient Tracking via Dynamic Programming}\label{section_Optimization}
The regularized tracking problem in~(\ref{P_regularized}) can be solved efficiently via dynamic programming. First, we recursively compute an \textit{accumulated regularized maximum energy map} $\mathbf{G}\in \mathbb{R}_+^{M\times N}$ column by column for all entries $(m,n)$ as follows
\begin{equation}
\mathbf{G}(m,n)=
\begin{cases}
\mathbf{Z}(m,n) + \lambda\log P_m, & n= 1;\\
\underset{m'}{\max} \{\mathbf{G}(m',n-1)+\lambda\log P_{m' m}\}&\\
\quad+ \mathbf{Z}(m,n), & n> 1. 
\end{cases}\label{DP_step1}
\end{equation}

After completing the calculation at column $n=N$, the maximum value of the $N$th column is denoted as $\hat{f}(N)$. Second, we find the optimal solution by backtracking from the maximum entry of the last column of the accumulated map $\mathbf{G}$. Specifically, we iterate $n$ from $N-1$ back to $1$ to solve for $\hat{f}(n)$ as follows
\begin{equation}
\begin{aligned}
& \hat{f}(n)=\underset{f(n)}{\text{argmax}}
& & \mathbf{G}(f(n),n)+\lambda \log P_{f(n) \hat{f}(n+1)}.\\
\end{aligned} \label{DP_step2}
\end{equation}
Note that we can avoid transitions from state $m'$ to state $m$ by setting $P_{m'm}=0$, since the regularization term would penalize the total energy to $-\infty$. If we assume uniform random walk transitions within the window containing $2k+1$ frequency bins around $f(n-1)=m'$, \textit{i.e.}, $P_{m'm}=\frac{1}{2k+1}$, $|m'-m|\leq k$, then problem~(\ref{P_regularized}) is degenerated, where the value $\lambda$ does not affect the solution.

\subsection{Trace Presence Test}\label{subsection_TraceExistenceDetection}
To determine the presence of a frequency component in a specific time interval, we first make independent decisions for every frame within the time interval on the presence of the frequency component, and then refine the decisions by considering neighborhood correlations. Adopting the terminology from the speech analysis, we refer to those frames with a frequency component as \textit{voiced} frames, or otherwise as \textit{unvoiced} frames. We propose to test the presence of a frequency component by evaluating the relative energy of the detected trace. A test statistic named the \textit{Relative Energy Ratio} (RER) is defined as follows:

\begin{equation}
\begin{aligned}
\text{RER}(n) = \frac{|\mathcal{F}(n)|\cdot\mathbf{Z}(\hat{f}(n), n)}{\sum_{m\in \mathcal{F}(n)}\mathbf{Z}(m, n)},
\end{aligned}\label{eq_RER}
\end{equation}
where $\mathcal{F}(n)\triangleq [1,M]\backslash [\max(1, \hat{f}(n)-\delta_f), \min(M,\hat{f}(n)+\delta_f)]$ is a conservative set of frequency indices that does not contain the frequency indices around the estimated frequency; $\delta_f$ is a predetermined parameter, and $|\cdot|$ is the cardinality of a set. It is evident that the higher RER$(n)$ is, the more probable that $n$th frame is voiced. The decision is made by comparing the test statistic $\text{RER}(n)$ with an empirically determined threshold $\Delta_{\text{RER}}$. A discussion about the optimal selection of $\Delta_{\text{RER}}$ will be presented later in Section~\ref{sssec::trace_detection}.

Next, we improve the initial frame-based presence detection results by merging nearby segments of the same type.
Specifically, we propose to merge two consecutive voiced segments if they are separated by an unvoiced segment shorter than $\Delta_1$, and then merge two unvoiced segments if they are separated by a voiced segment shorter than $\Delta_2$. Here, $\Delta_1$ and $\Delta_2$ are the upper bounds for determining unvoiced and voiced segments that allow merging, respectively.
Fig.~\ref{fig_OfflineAMTCExample01}(e) and Fig.~\ref{fig_OfflineAMTCExample01}(f) illustrate two examples of the decision making process in which the dash--dot blue curve corresponds to the initial decision results and the solid blue curve corresponds to the refined/final decision results. 
Note that the final decisions have successfully excluded short segments given by the initial decisions.

\begin{algorithm}[!t]
\caption{Offline Adaptive Multi-Trace Carving (offline-AMTC)}\label{algo_AMTC}
\begin{algorithmic}[1]
\Procedure{AMTC}{$\mathbf{Z},L$}\Comment{$L$: number of output traces}
   \State $\mathbf{Z}_{(1)}\gets \mathbf{Z}$
   \State $\hat{\mathbf{f}}_{(1)}\gets \underset{\mathbf{f}}{\text{argmax}}
\,E_{\mathbf{Z}_{(1)}}(\mathbf{f})+\lambda P(\mathbf{f})$
   \State $\hat{\mathbf{v}}_{(1)}\gets \text{DetectPresence}(\mathbf{Z}_{(1)}, \hat{\mathbf{f}}_{(1)}, \Delta_{\text{RER}}, \Delta_1, \Delta_2)$\footnotemark
   \For{$l \gets 2$ to $L$} 
      \State Update $\mathbf{Z}_{(l)}$ according to~(\ref{eq_ZCompensation})
      \State $\hat{\mathbf{f}}_{(l)}\gets \underset{\mathbf{f}}{\text{argmax}}
\,E_{\mathbf{Z}_{(l)}}(\mathbf{f})+\lambda P(\mathbf{f})$
	  \State $\hat{\mathbf{v}}_{(l)}\gets \text{DetectPresence}(\mathbf{Z}_{(l)}, \hat{\mathbf{f}}_{(l)}, \Delta_{\text{RER}}, \Delta_1, \Delta_2)$
   \EndFor 
   \State \textbf{return} $\hat{\mathbf{f}}_{(1:L)}$, $\hat{\mathbf{v}}_{(1:L)}$
\EndProcedure
\end{algorithmic}
\end{algorithm}
\footnotetext{DetectPresence$(\cdot)$ refers to the trace presence detection algorithm described in Section~\ref{subsection_TraceExistenceDetection}. $\hat{\mathbf{v}}_{(l)}\in \{0,1\}^N$ is the trace presence decision with $0$ as unvoiced and $1$ as voiced.}

\section{Tracking Multiple Traces via Iterative Frequency Compensation}\label{section_AMTC}

In the previous section, we have introduced a single frequency trace tracking and detection method using dynamic programming and trace presence testing, respectively. For some forensic tasks such as extracting pulse rate from the face video containing subject's motion, as shown in Fig.~\ref{subfig_ExpHROri}, the presence of multiple traces in the frequency range of interest is possible, and the dominating trace in the spectrogram might not be the one of interest. A crude deployment of any single-trace tracking method on such tasks would generate completely wrong answers. To address this problem, we extend the single-trace tracking method to be able to track multiple traces by extracting each trace iteratively to find all candidates. We name this method the Adaptive Multi-Trace Carving (AMTC). In the rest part of this section, we first present the offline version of AMTC (offline-AMTC), when the trace estimate is optimized according to the entire available signal. We next adapt the offline-AMTC to an efficient online version (online-AMTC), which runs in near real time with low delay.
\subsection{Offline-AMTC}\label{subsection_offlineAMTC}

\begin{figure}[t]
\centering
\includegraphics[width=3.2in]{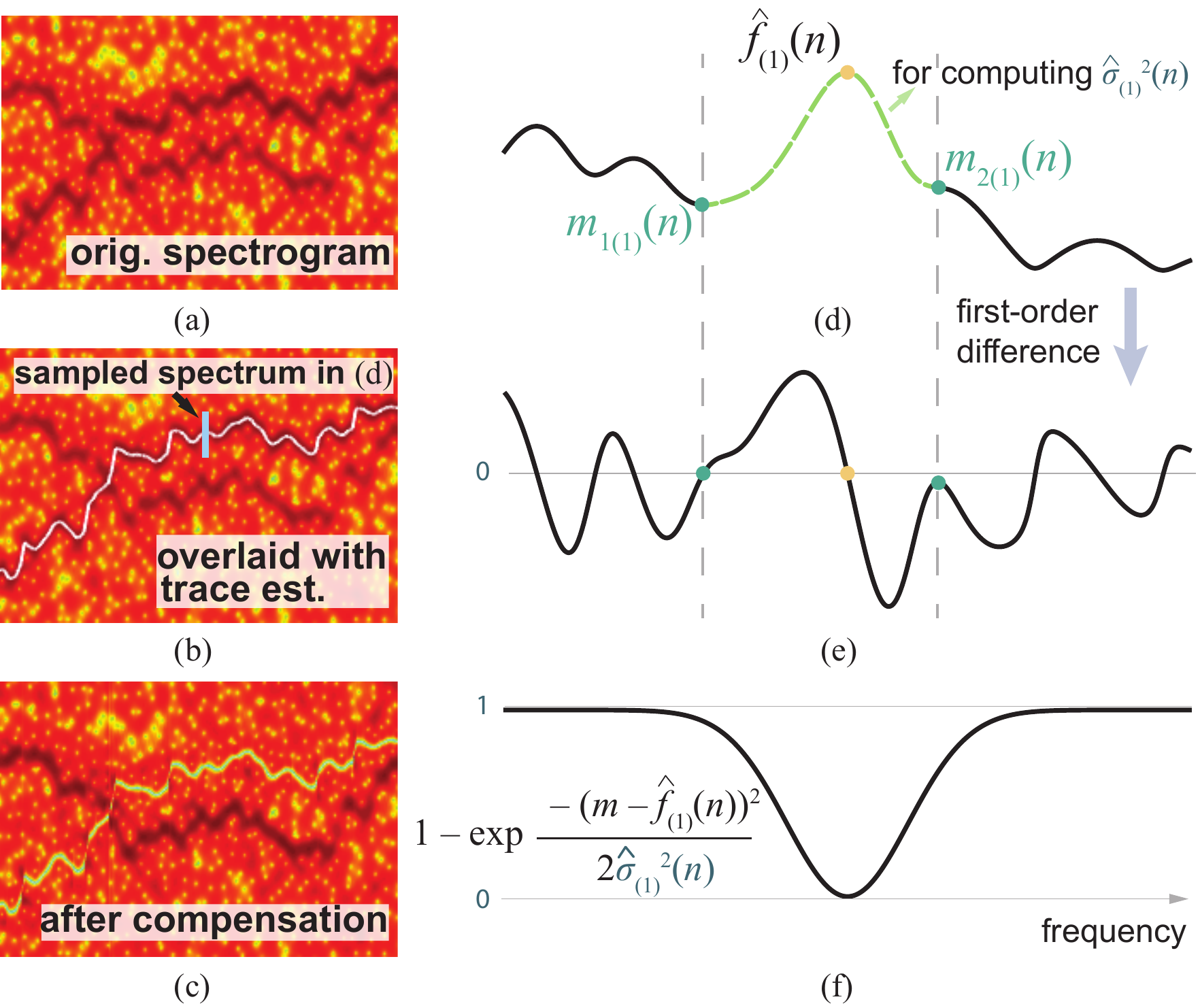}%
\caption{Illustrations for the trace compensation process: (a) the spectrogram of a synthetic $-8$~dB signal with two frequency components; (b) first trace estimate by AMTC (white line); (c) the spectrogram after the first trace compensation; (d) sampled spectral distribution centered at $\hat{f}_{(1)}(n)$, where $n = 400$, see the vertical line in (b); (e) the first-order difference of the spectral function in (d); (f) the generated point-wise compensation weights. The value of $\hat{\sigma}_{(1)}^2$ is determined by the values within the green dashed segment in (d).}\label{fig::trace_comp_details}
\end{figure}

Similar to the iterative nature of the seam carving algorithm~\cite{avidan2007seam}, a greedy algorithm can be used to search multiple traces by iteratively running the single-trace tracker proposed in Section~\ref{section_Problem}.
Since the single-trace tracker only extracts the dominating trace from the spectrogram, the previously detected traces need to be erased or compensated before invoking the single-trace tracker in the next iteration.

Below, we describe the trace compensation process.
Suppose $\hat{\mathbf{f}}_{(l)}$ is the estimated frequency trace at the $l$th iteration. For each time frame of the spectrogram, \textit{i.e.}, $\mathbf{Z}_{(l)}(1:M, n)$, we search for a left boundary point $m_{1(l)}(n)$ from $\hat{f}_{(l)}(n)$ to its left side. We set $m_{1(l)}(n) = m$, where $m$ is the nearest point to $\hat{f}_{(l)}(n)$ that is either a local minimum point in $\mathbf{Z}_{(l)}(1:M, n)$ or a local minimum point in the first-order difference of $\mathbf{Z}_{(l)}(1:M, n)$. The search of the right boundary point $m_{2(l)}(n)$ works similarly except it considers the local maximum point in the first-order difference of $\mathbf{Z}_{(l)}(1:M,n)$. In this paper, we call $\mathbf{Z}_{(l)}(m_{1(l)}(n): m_{2(l)}(n), n)$ \textit{the effective peak} of $\hat{f}_{(l)}(n)$.

One example of the trace compensation process is shown in Fig.~\ref{fig::trace_comp_details}. The plot in (d) shows the spectral energy distribution centered at $\hat{f}_{(1)}(n)$, which corresponds to the light blue vertical line in (b). In this case, $m_{1(1)}(n)$ is selected as the first local minimum point, and $m_{2(1)}(n)$ as the local maximum point in the first-order difference of $\mathbf{Z}_{(l)}(1:M, n)$. Based on $m_{1(l)}(n)$ and $m_{2(l)}(n)$, we propose to use a flipped Gaussian-shaped function to compensate the energy of the estimated frequency component. The compensated power spectrum at the $(l+1)$st iteration is updated point-wise in $n$ and $m$ as follows
\begin{equation}
\begin{aligned}
\mathbf{Z}_{(l+1)}(m,n) &\leftarrow \left[1-\exp \frac{-\left(m-\hat{f}_{(l)}(n)\right)^2}{2\hat{\sigma}^2_{(l)}(n)}\right]\cdot \mathbf{Z}_{(l)}(m,n),\\
\hat{\sigma}^2_{(l)}(n) &= \frac{\sum_{m=m_{1(l)}(n)}^{m_{2(l)}(n)} \mathbf{Z}_{(l)}(m,n)(m-\hat{f}_{(l)}(n))^2}{\sum_{m=m_{1(l)}(n)}^{m_{2(l)}(n)}\mathbf{Z}_{(l)}(m,n)}
\end{aligned}
\label{eq_ZCompensation}
\end{equation}
where $\hat{\sigma}^2_{(l)}(n)$ is used to quantify the width of the effective peak at the $l$th iteration. The pseudo code of the offline-AMTC is shown in Algorithm~\ref{algo_AMTC}. In Fig.~\ref{fig_OfflineAMTCExample01}, we give an example of two-trace estimation process on a synthetic heart beat signal. The final estimate is almost identical with the ground truth, and the unvoiced segments are successfully detected.
\par If we define $L$ as the number of traces to track, the computational complexity for the offline-AMTC is $O(NLM^2)$. To compare, the fHMM methods \cite{wohlmayr2011probabilistic,liu2017speaker} requires $O(NLM^{L+1})$ without considering operations for computing emission probability. The efficiency of offline-AMTC is mostly explained by the idea of the introduced iterative search. We will later show in Section~\ref{section_exp} that the demonstrated efficiency is not achieved at the expense of performance drop.

\begin{algorithm*}
\caption{Online-AMTC at time $n$}\label{algo_onlineAMTC}
\begin{algorithmic}[1]
\Procedure{AMTC}{$\mathbf{Z}_{(1:L)}(\tau_1:\tau_2-1)$, $\mathbf{G}_{(1:L)}(\tau_1:\tau_2-1)$,  
$\hat{f}_{(1:L)}^{\textrm{pre}}(\tau_1:\tau_2-1)$, $\mathbf{Z}_{(1)}(\tau_2)$}       \Comment{$\tau_1\triangleq n-k_1$, $\tau_2\triangleq n+k_2$.}
	\State $\mathbf{Z}_{(1)}(\tau_1:\tau_2)\gets$ concatenate $\mathbf{Z}_{(1)}(\tau_1:\tau_2-1)$ and $\mathbf{Z}_{(1)}(\tau_2)$; $T_e \gets \tau_2-1$
    \State Update $\mathbf{G}_{(1)}(\tau_2)$ according to~(\ref{DP_step1}) using $\mathbf{G}_{(1)}(\tau_2-1)$ and $\mathbf{Z}_{(1)}(\tau_2)$

    \For {$l \gets 1$ to $L$}
    	\State Estimate $\hat{f}_{(l)}(T_e+1:\tau_2)$ according to (\ref{DP_step2}) using $\mathbf{G}_{(l)}(T_e+1:\tau_2)$
        \If {$l<L$}
        	\State Update $\mathbf{Z}_{(l+1)}(T_e+1:\tau_2)$ according to (\ref{eq_ZCompensation}) using $\mathbf{Z}_{(l)}(T_e+1:\tau_2)$ and $\hat{f}_{(l)}(T_e+1:\tau_2)$
        \EndIf
        \For {$i \gets T_e$ to $\tau_1$}
        	\State Estimate $\hat{f}_{(l)}(i)$ according to (\ref{DP_step2}) using $\hat{f}_{(l)}(i+1)$ and $\mathbf{G}_{(l)}(i)$
            \If {$l<L$}
            	\State Update $\mathbf{Z}_{(l+1)}(i)$ according to (\ref{eq_ZCompensation}) using $\mathbf{Z}_{(l)}(i)$ and $\hat{f}_{(l)}(i)$
            \EndIf
            \If {$\hat{f}_{(l)}(i)==\hat{f}_{(l)}^{\textrm{pre}}(i)$}
            	\State Update $\mathbf{G}_{(l+1)}(i+1:\tau_2)$ according to (\ref{DP_step1}) using $\mathbf{Z}_{(l+1)}(i+1:\tau_2)$ and $\mathbf{G}_{(l+1)}(i)$
                \State $\hat{f}_{(l)}(\tau_1:i) \gets \hat{f}^{\textrm{pre}}_{(l)}(\tau_1:i)$;  $T_e\gets i$;  \textbf{break}
                
            \ElsIf{$i==\tau_1$}
            	\State Update $\mathbf{G}_{(l+1)}(\tau_1:\tau_2)$ according to (\ref{DP_step1}) using $\mathbf{Z}_{(l+1)}(i :\tau_2)$;  $T_e \gets i$
            \EndIf
         \EndFor
        \State $\hat{v}_{(l)}(\tau_1:\tau_2)=\text{DetectPresence}(\mathbf{Z}_{(l)}(\tau_1:\tau_2), \hat{f}_{(l)}(\tau_1:\tau_2), \Delta_{\text{RER}}, \Delta_1, \Delta_2)$   \EndFor
    
    \State \textbf{return} $\hat{f}_{(1:L)}(n)$, $\hat{v}_{(1:L)}(n)$, $\mathbf{Z}_{(1:L)}(\tau_1+1:\tau_2)$, $\mathbf{G}_{(1:L)}(\tau_1+1:\tau_2)$, $\hat{f}_{(1:L)}(\tau_1+1:\tau_2)$ \EndProcedure
\end{algorithmic}
\end{algorithm*}

\begin{figure}
\centering
\includegraphics[width=3.5in]{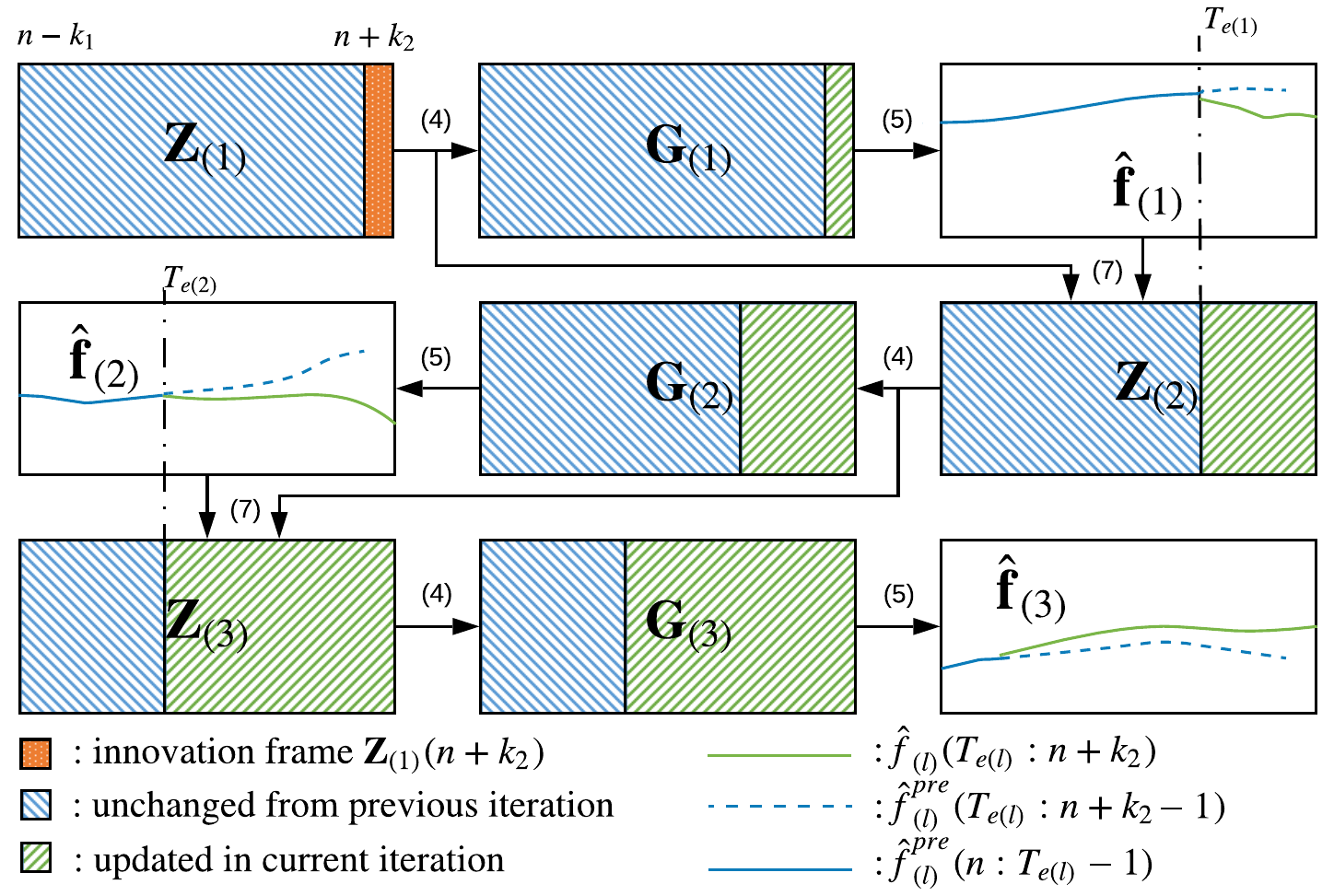}%
\caption{A flowchart for the online-AMTC algorithm for three-trace estimation process at $t$th iteration. $(\cdot)$ above arrows indicates the index of the equation being used.}\label{fig_AMTC_online_diagram}
\end{figure}

\subsection{Online-AMTC: An Efficient and Near Real-Time Approach}\label{subsection_onlineAMTC}
For the use cases of real-time tracking, we propose \emph{online-AMTC} by adapting the offline-AMTC to track multiple frequency components in near real time (NRT) with a delay of $k_2$ time units.
In this NRT scenario, the tracking goal at the time instant $n$ is to estimate $f_{(1:L)}(n)$ based on the available spectrogram information $\mathbf{Z}_{(1)}(1:n+k_2)$\footnote{For concise representation, we use $\mathbf{G}(n_1:n_2)$ and $\mathbf{Z}(n_1:n_2)$ as a shorthand for $\mathbf{G}(1:M, n_1:n_2)$ and $\mathbf{Z}(1:M, n_1:n_2)$, respectively.}.

A naive \emph{brute-force approach} can be constructed by iteratively running the offline-AMTC from time instant $1$ to $n+k_\rev{2}$. For each time instant $n$, the time complexity is $O(nLM^2)$. If the total number of the spectral frame\rev{s} is $N$, this slow but accurate approach will require $O(NM)$ in space and $\sum_{n = 1} ^{N}O(nLM^2) = O(N^2LM^2)$ in time. The computational complexity in space and time increase\rev{s} linearly and quadratically in $N$, respectively. This unbounded increase will eventually lead to both memory overflow and CPU overload, especially when systems are deployed for a longer term.

Another way to adapt the offline-AMTC for the NRT scenario and to solve the computational issue of the brute-force approach is to segment the time signal into non-overlapping \rev{chunks}, and let the offline-AMTC run independently on each segment. 
This segment-based approach (aka, \textit{the segment-based offline-AMTC}) trades accuracy for space and time complexity since the information before the block of interest is not used in the estimation process. This issue is more severe for those with shorter block length, as confirmed by the experimental results in Section~\ref{sssec::single_trace}.

The online-AMTC is developed to address the storage and computational issues mentioned above with comparable accuracy with the offline-AMTC. We propose to use a fixed-length queue buffer for storing and updating the intermediate values of $\mathbf{Z}$, $\mathbf{G}$, and $\hat{f}$. As a result, the running time and the memory requirement are greatly reduced and are independent of time $n$. 
 
\par We introduce the online-AMTC by first discussing online iterations for the estimation process of the first trace. The processing flow of the online-AMTC algorithm at the instant $n$ is illustrated in Fig.~\ref{fig_AMTC_online_diagram}. Suppose the allowed delay length is $k_2$, and $\hat{f}_{(1)}(n-1)$ has been computed by backtracking from the accumulated regularized maximum energy map $\mathbf{G}_{(1)}(n-1:n+k_2-1)$. At the arrival of the next innovation frame $\mathbf{Z}_{(1)}(n+k_2)$ (the orange frame in Fig.~\ref{fig_AMTC_online_diagram}), our goal is to estimate $\hat{f}_{(1)}(n)$. From the forward update rule of $\mathbf{G}$ in~(\ref{DP_step1}), it can be seen that $\mathbf{G}_{(1)}(n:n+k_2-1)$ would remain unchanged compared to the output in the previous time instant $n-1$. We therefore only need to update the right most frame $\mathbf{G}_{(1)}(n+k_2)$ given $\mathbf{G}_{(1)}(n+k_2-1)$ and the innovation frame $\mathbf{Z}_{(1)}(n+k_2)$ as shown in the middle box of the first row of Fig~\ref{fig_AMTC_online_diagram}. We can then obtain $\hat{f}_{(1)}(n)$ via backtracking from $\mathbf{G}_{1}(n:n+k_2)$ according to~(\ref{DP_step2}). 

Denote the previous estimates at time $n-1$ obtained from the backtracking process that leads to $\hat{f}_{(1)}(n-1)$ as $\hat{f}^{\textrm{pre}}_{(1)}(n-1:n+k_2-1)$. During the backtracking process for $\hat{f}_{(1)}(n)$, if $\hat{f}_{(1)} = \hat{f}_{(1)}^{\text{pre}}$ at the time instant index $T_e\in [n, n+k_2)$, $\hat{f}_{(1)}(n:T_e)$ stays unchanged as $\hat{f}_{(1)}^{\text{pre}}(n:T_e)$. This claim holds because $\mathbf{G}_{(1)}(n:T_e)$ remains the same during the process. In this regard, we consider storing and updating $\hat{f}^{\textrm{pre}}_{(1)}(n-1:n+k_2-1)$ in a buffer, whereby the update process of $\hat{f}^{\textrm{pre}}_{(1)}$ stops at the instant $T_e$ if $\hat{f}_{(1)}(T_{e})= \hat{f}^{\textrm{pre}}_{(1)}(T_e)$, as shown in the right box of the first row of Fig.~\ref{fig_AMTC_online_diagram}. In this way, the computation complexity is further reduced.

\par Different from the estimation process for the first trace, any change from previous trace estimation $\hat{f}_{(1:l-1)}$ would have influence on the formation of $\mathbf{Z}_{(l)}$, $\mathbf{G}_{(l)}$, and therefore $\hat{f}_{(l)}$. In order to obtain a robust estimate for $\hat{f}_{(l)}$, $l>1$, we introduce a look-back length, $k_1>0$ in this process. As demonstrated from second and third rows in Fig.~\ref{fig_AMTC_online_diagram}, for $l$th trace estimation at time instant $n$, we utilize the previous trace estimates $\hat{f}_{(l-1)}(n-k_1:n+k_2)$ and $\mathbf{Z}_{(l-1)}(n-k_1:n+k_2)$ to obtain new $\mathbf{Z}_{(l)}(n-k_1:n+k_2)$ and $\mathbf{G}_{(l)}(n-k_1:n+k_2)$, and thus $\hat{f}_{(l)}(n-k_1:n+k_2)$. Efficient backtracking can also be achieved using the previous backtracking strategy, same as the case in estimating the first trace. The details of the online-AMTC algorithm at the $n$th iteration is shown in Algorithm~\ref{algo_onlineAMTC}.

\par The worst-case computational complexity for the online-AMTC is $O(N(k_1+k_2)LM^2)$, which appears to be $(k_1+k_2)$ times higher than the offline version. In a statistical sense, however, the expected complexity of the online-AMTC is much less than the worst-case analysis result because the probability that an entire trace estimate being changed from the previous one is low at each time instant. We will compare the average computation time running the offline- and online-AMTC in Section~\ref{sssec::multitraceExp}.

\begin{figure}
\centering
\includegraphics[width=3.5in]{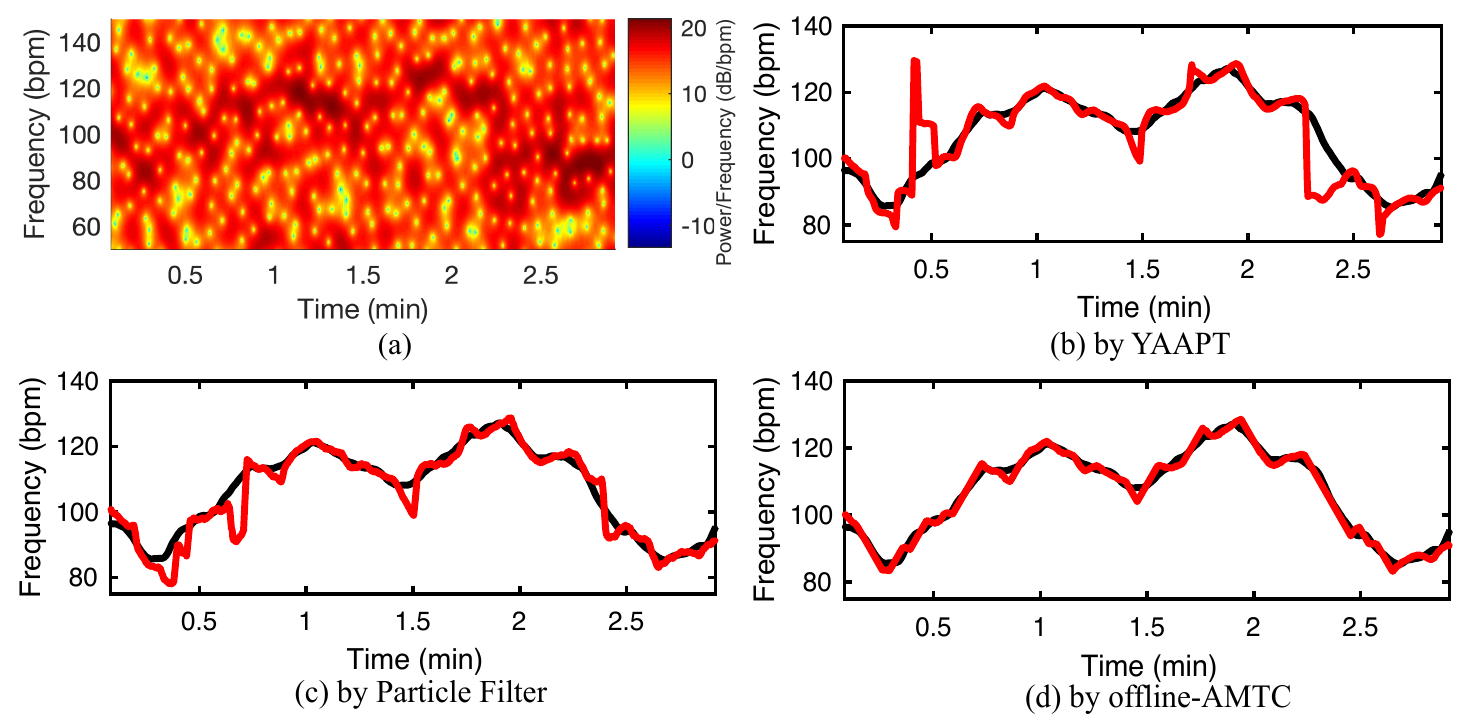}%
\caption{(a) Spectrogram of a synthetic $-10$~dB signal with one frequency component; Trace tracking results (red line) by (b) YAAPT, (c) particle filter, and (d) offline-AMTC, respectively. The ground-truth trace is shown in dashed black line in each plot.}\label{fig_singletrace_example}
\end{figure}

\begin{figure*}
\centering
\subfigure{\includegraphics[width=2.2in]{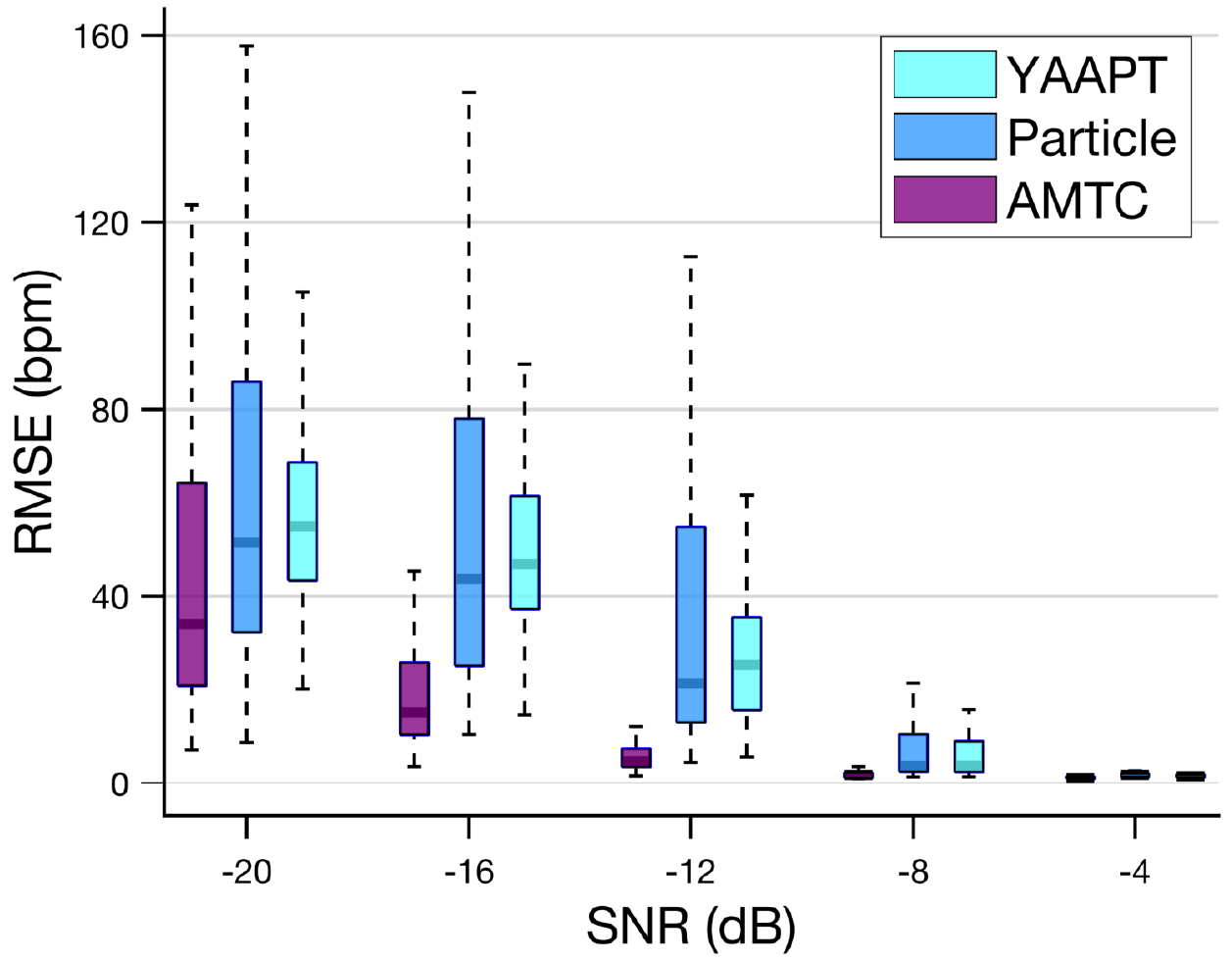}%
\label{fig_first_case}}
\hspace{1em}
\subfigure{\includegraphics[width=2.2in]{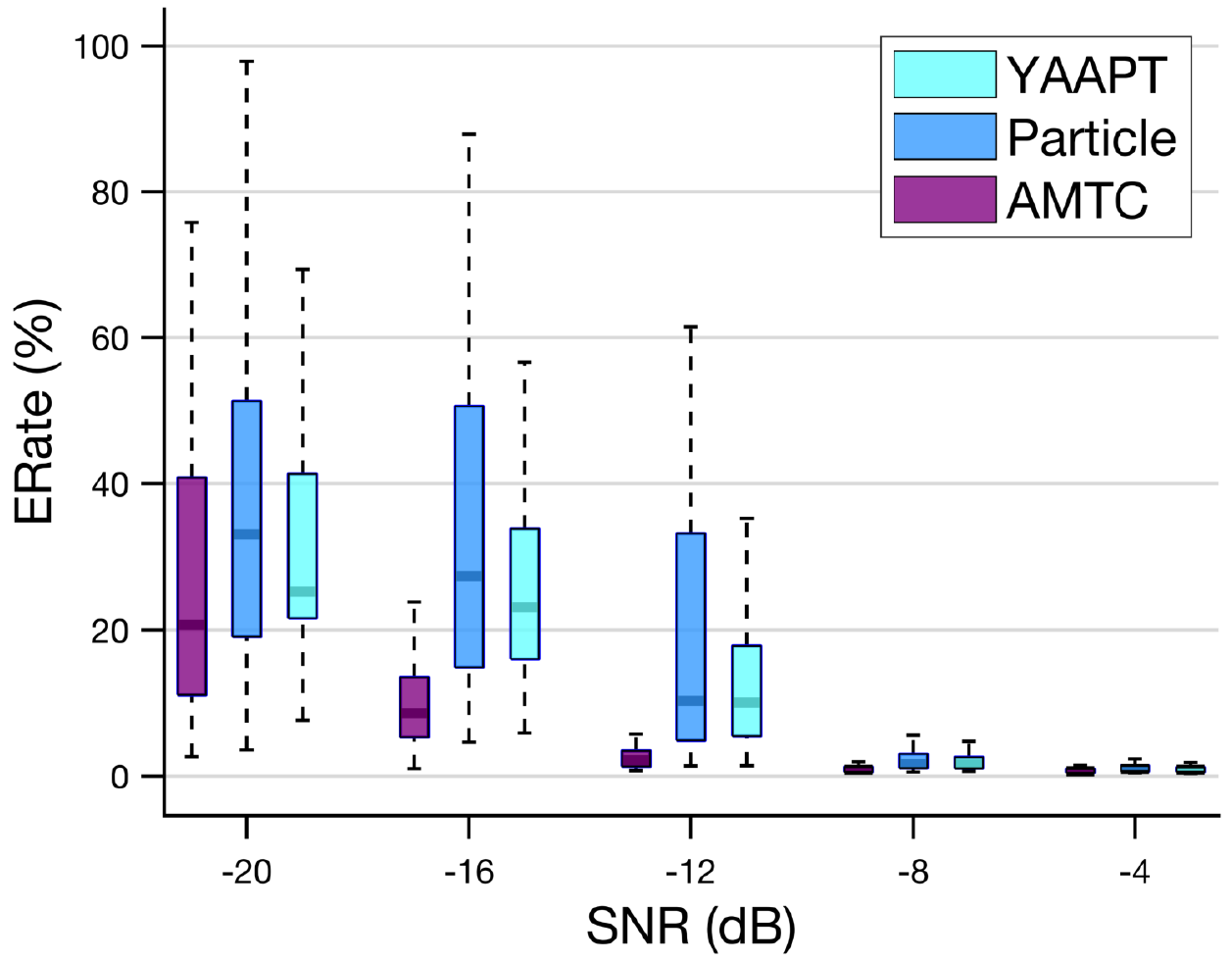}%
\label{fig_second_case}}
\hspace{1em}
\subfigure{\includegraphics[width=2.2in]{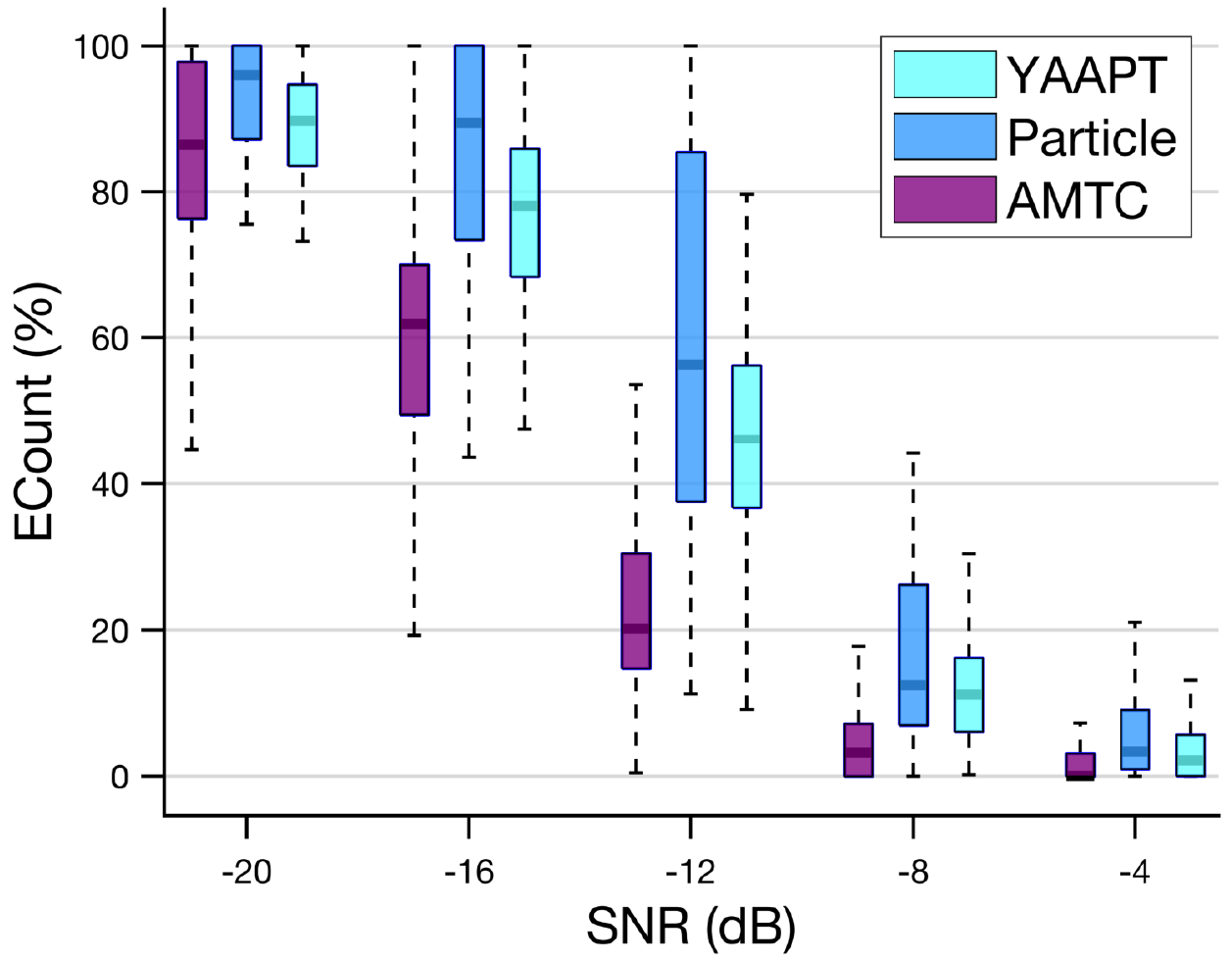}%
\label{fig_third_case}}
\subfigure{\includegraphics[width=2.2in]{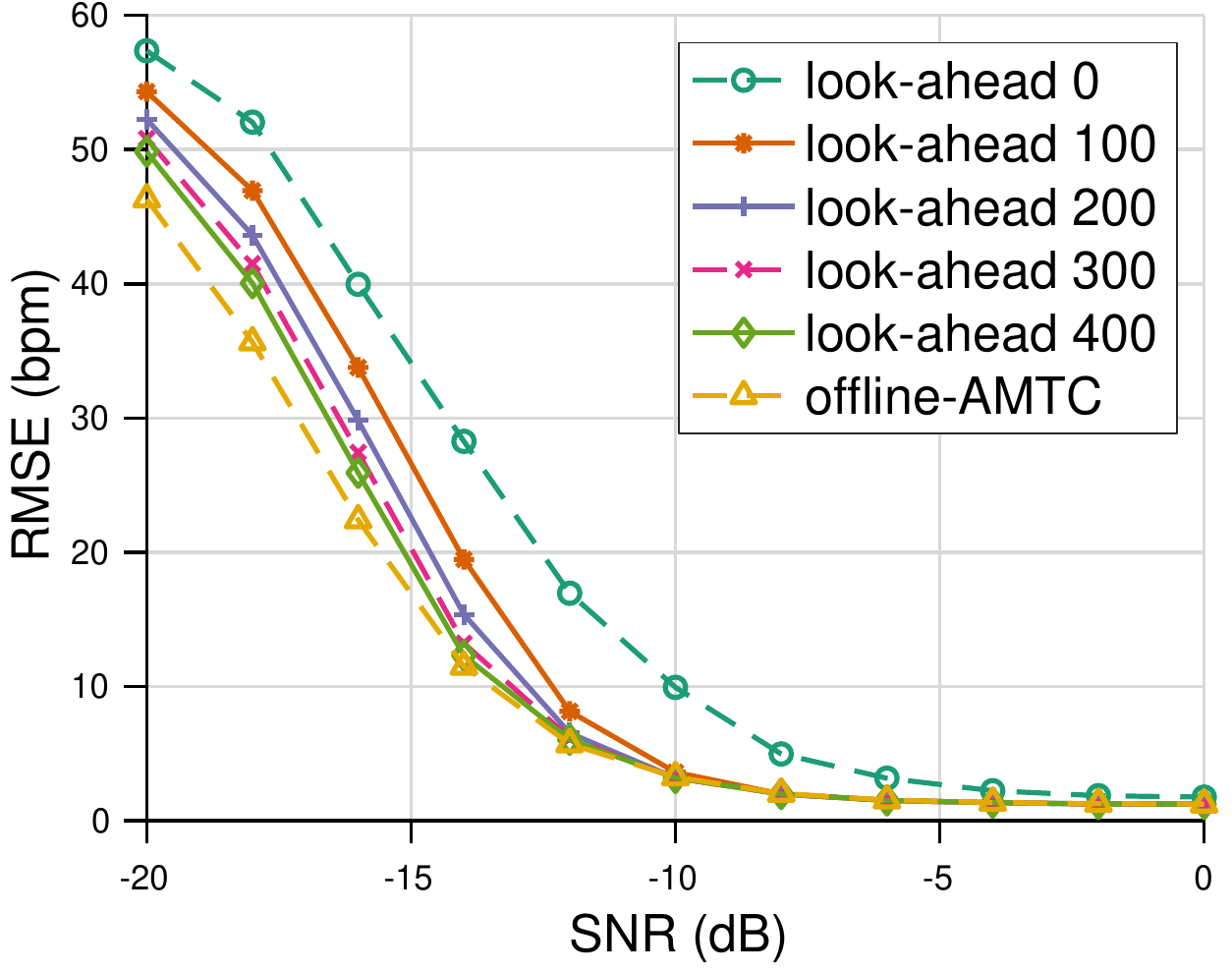}%
\label{fig_fourth_case}}
\hspace{1em}
\subfigure{\includegraphics[width=2.2in]{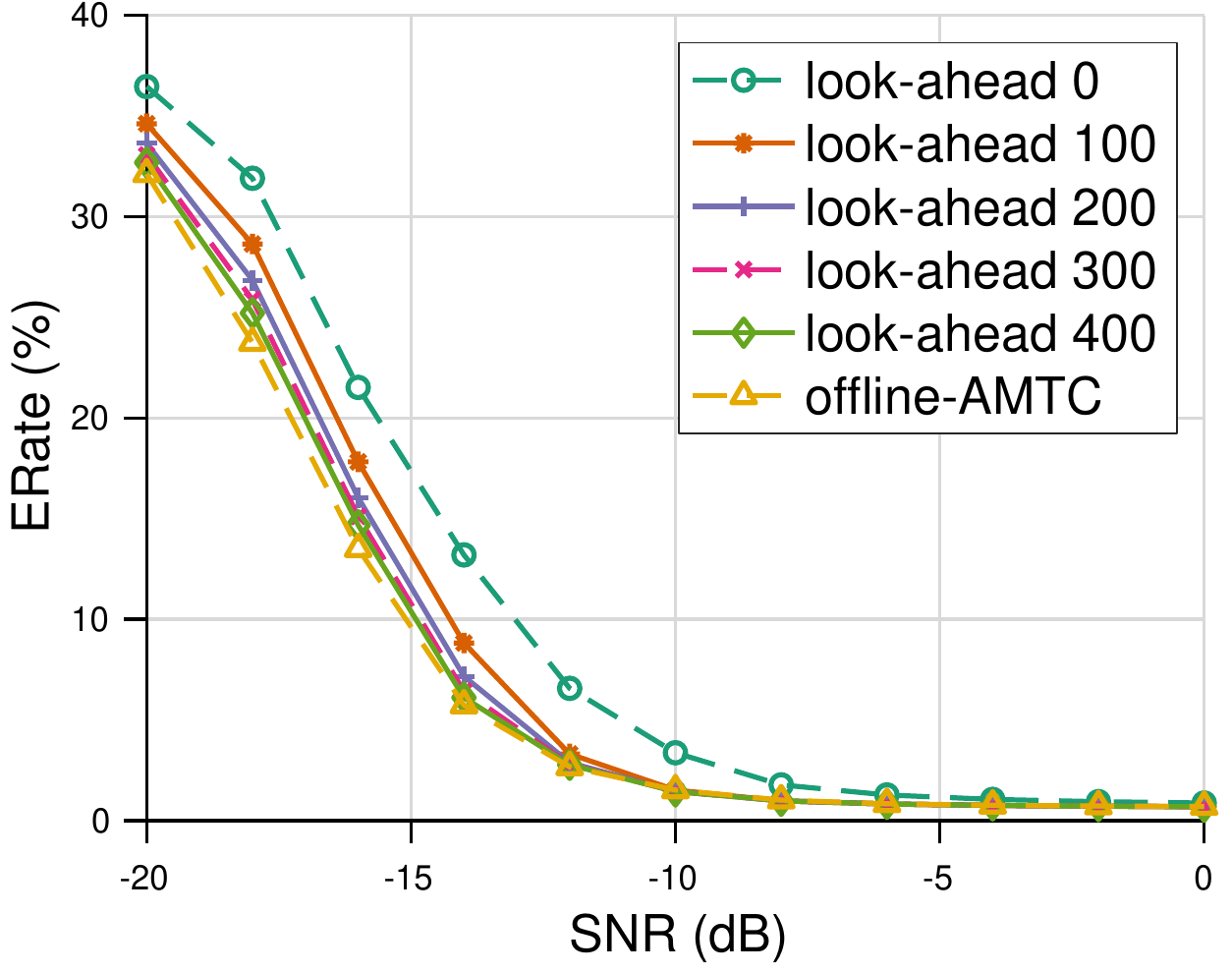}%
\label{fig_fifth_case}}
\hspace{1em}
\subfigure{\includegraphics[width=2.2in]{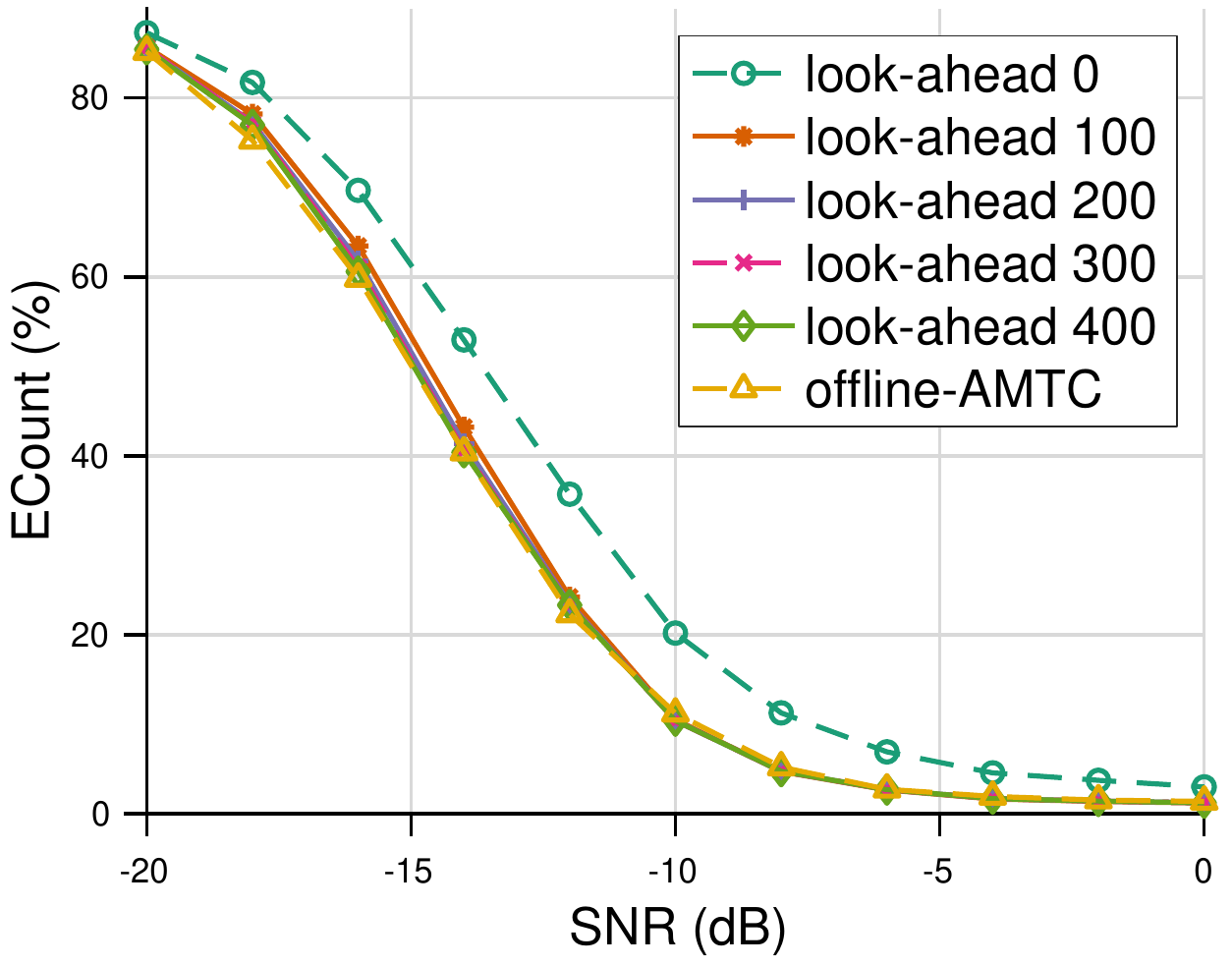}%
\label{fig_sixth_case}}
\caption{First row: Comparison of the performance of single-trace tracking by the proposed offline-AMTC, particle filter, and YAAPT methods at different levels of SNR. Statistics of the \textsc{Rmse}, the \textsc{ERate}, and \textsc{ECount} of frequency estimates are summarized using box plots. Second row: Trace tracking performance by the online-AMTC with different levels of look-ahead window length and SNR. The results for the offline-AMTC are also shown in the plots for the comparison purpose.}
\label{fig_sim}
\end{figure*}

\section{Experiments and Performance Analysis}\label{section_exp}

In this section, we carry out experiments to examine the performances of the AMTC algorithms, including the baseline version and multiple variants. We first use synthetic data with known ground truth, and then study with real-world data.
\subsection{Simulation Results and Comparison with Known Ground Truth}\label{section_exp_subsection_simu}
\subsubsection{Single Trace}~\label{sssec::single_trace}
We first evaluate the performance of the AMTC algorithm using synthetic data containing a single frequency trace. The AMTC algorithm is compared with the Particle Filter~\cite{shi2003spectrogram} and the local peak based YAAPT~\cite{kasi2002yet} methods. For the offline-AMTC and the online-ATMC, we used the uniform random walk model specified in Section~\ref{section_Optimization}. The hyperparameters $k$, $\Delta_{\text{RER}}$, $\Delta_1$, $\Delta_2$, $k_1$, and $k_2$ were set as $3$, $2.41$, $30$, $30$, $50$, and $100$ throughout the paper, respectively, unless otherwise stated. For the Particle Filter method, the number of particles was set to 1024. For each test signal, we generated a time-varying pulse rate trace present from the beginning to the end of the timeline. More specifically, denote $s[n]$ as the temporal measurement of the corrupted frequency signal, $s[n] = \sin \Phi[n]+\epsilon[n]$, where $\Phi[n] = \Phi[n-1]+2\pi f[n]/f_s$, $f[n]$ is the time-varying synthesis frequency, $f_s$ is the sampling rate set to 30 $\text{Hz}$, and $\epsilon[n]$ is the noise quantified by a zero-mean white Gaussian process. The variance of $\epsilon[n]$ is an adjustable parameter for achieving different SNR levels. To generate frequency signals $f[n]$ that behave similarly as real-world pulse rate signals, we trained a 9-tap autoregressive model using heart rate signals collected by a Polar H7 chest belt in both exercise mode and still mode. We use beat per minute (bpm) as the frequency unit. The duration of each test signal is 3 minutes. The spectrograms were generated by short-time Fourier transform (STFT) with a rectangular window of length $10$ seconds and $98\%$ overlap between adjacent frames. We padded zeros to the end of each frame to make neighboring frequency bins $0.17$ bpm apart. 
\par We generated 500 trials under each of the five SNR conditions, or 250 for each mode (namely, the exercise and the still cases) using the estimated parameters of the autoregressive models. We used three metrics. Namely, the root mean squared error (\textsc{Rmse}), the error rate (\textsc{ERate}), and the error count (\textsc{ECount}) defined as follows to evaluate the performance:
\begin{itemize}
\item \textsc{Rmse} $= \sqrt{\frac{1}{T} \sum_{t=1}^T(\hat{f}_t-f_t)^2}$\,,
\item \textsc{ERate} $= \frac{1}{T} \sum_{t=1}^T \left| \hat{f}_t-f_t \right| \Big/ f_t$\,,
\item \textsc{ECount} $= \left| \{t\in [1, T] : \left|\hat{f}_t-f_t\right|\Big/ f_t >\tau\} \right|\Big/ T$\,,
\end{itemize}
where $|\{\cdot\}|$ denotes the cardinality of a countable set, $\hat{f}_t$ and $f_t$ are the frequency estimate and the ground-truth frequency at $t$th time frame respectively, and $\tau$ was chosen to be $0.03$ empirically determined from the spread of the frequency components. Fig.~\ref{fig_singletrace_example} shows tracking results of a $-10$~dB synthetic signal with one frequency component using 
offline-AMTC, YAAPT, particle filter, respectively. In this example, AMTC outputs the best trace estimate among the three without much deviation from the ground truth. The results of overall performance are shown in the first row of Fig.~\ref{fig_sim} in terms of box plots that each box compactly shows the median, upper and lower quantiles, and the max and min values of a dataset. It is evident from the box plots that, under all SNR levels, offline-AMTC generally outperforms the particle filter method and the YAAPT not only in terms of the average but also in the variance of the error statistics.

\par Next, we tested the online-AMTC algorithm using different look-ahead \rev{($k_2$)} time lengths. The evaluation was conducted using the same setting mentioned above, and the averaged behavior of each look-ahead length is plotted in the second row of Fig.~\ref{fig_sim}. The numbers in the legends indicate the lengths of look-ahead window lengths represented by the number of time bins in the spectrogram. We have two observations from these plots. First, a performance jump from no look-ahead versus 100-bin look-ahead length is observed.
The performance curve
moves closer to that of the offline-AMTC as the look-ahead
length increases.
This observation coincides with the intuition that a small look-ahead length would cause the online trace estimator to find a locally optimum solution. Second, given the shape of the curve, the performance starts to converge from $\text{SNR}=-10$~dB upwards and is almost identical when the look-ahead length is greater than $100$ frames. This trend of performance convergence is also expected as the signal quality is high enough for AMTC to track the correct trace. 

\begin{table}[h]
\centering
\ra{1.3}
\caption{Performance comparison between the online-AMTC, the brute-force approach, and the segment-based offline-AMTC}
\label{table::algo_comparison}
\centering
\begin{tabular}{@{}rrrrr@{}}\toprule
\multicolumn{2}{l}{\textbf{NRT Delay ($\#$ of Frames)}} & {\bfseries 0} & {\bfseries 50} & {\bfseries 100}\\

\midrule
\multicolumn{3}{l}{\textbf{Runtime (seconds per $100$ frames)}}\\
& Online-AMTC &                0.22 & 0.33 & 0.38\\
& Brute-Force &          1.59 & 2.24 & 2.50\\
& Segment-Base Offline-AMTC    & 0.09 & 0.09 & 0.09 \\

\midrule
\multicolumn{3}{l}{\textbf{\textsc{ERate} ($\%$)}}\\
& Online-AMTC                 & 6.99 &  3.36 &  3.26 \\
& Brute-Force           & 6.99 &  3.36 &  3.26 \\
& Segment-Based Offline-AMTC    & 23.32 & 16.12 & 12.01 \\
\bottomrule
\end{tabular}
\end{table}

\begin{figure}
\centering
\subfigure[]{\includegraphics[width=1.7in]{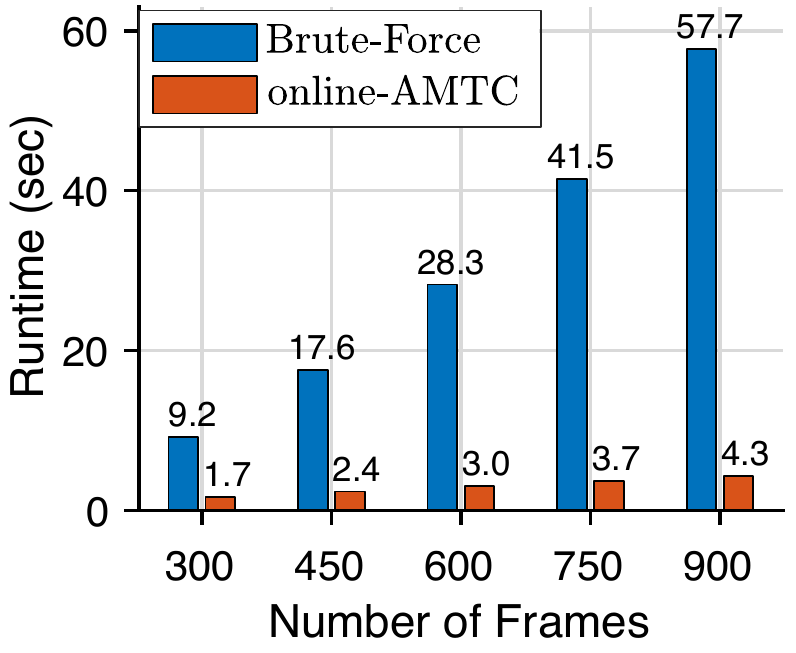}}
\subfigure[]{\includegraphics[width=1.7in]{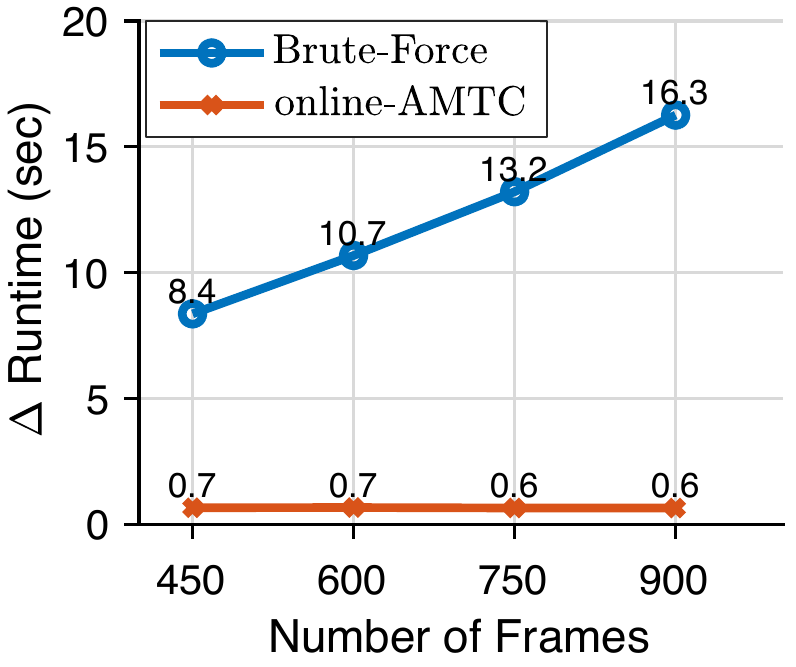}}
\caption{(a) The runtime function of the brute-force approach and the online-AMTC in near real-time scenario. The trace estimates are identical for the two methods. (b) \rev{F}irst-order difference function of the runtime curves in (a). }
\label{fig_timeOnlineVsBruteForce}
\end{figure}
Finally, we compared the performance \rev{among} the online-AMTC, the brute-force approach, and the segment-based offline-AMTC in terms of their runtime and \textsc{ERate} in NRT scenario using a 2014 MacBook Pro with a $2.3$~GHz Intel Core i5 processor. In Table~\ref{table::algo_comparison}, we show the performance comparison when the SNR is $-12$~dB with the experimental setting specified in this section. Note that the \textsc{ERate} of the segment-based offline-AMTC is more than three times higher than that of the online-AMTC when the NRT delay equals $100$ frames, and more than four times higher than that of the online-AMTC when the delay equals $50$ frames. The advantage in computational time complexity of the segment-based offline-AMTC is at a cost of a significant drop in estimation accuracy, where we observe an increase of \textsc{ERate} by more than $8.5\%$ when the NRT delay is up to $100$ frames. We also observed that the runtime of the brute-force approach is significantly higher than that of the online-AMTC in the same estimation accuracy. This can also be seen in Fig.~\ref{fig_timeOnlineVsBruteForce}(a), where we compare the runtime of the online-AMTC and the brute-force approach in terms of the number of processed frames with the look-ahead length equaling $100$ frames. We see that the online-AMTC runs ten times faster than the brute-force approach when the total frames to process is more than $750$. From the first-order difference function of the runtime curves shown in Fig.~\ref{fig_timeOnlineVsBruteForce}(b), the runtime of the brute-force approach and the online-AMTC appears to be quadratic and linear in the number of processed frames. This observation coincides with our analysis in Section~\ref{subsection_onlineAMTC}.

\begin{table*}
\centering
\ra{1.3}
\caption{Average Performance of fHMM and AMTC on multi-trace tracking test}
\label{table_multi_trace_err}
\centering
\begin{tabular}{@{}rrrrrrrrrr@{}}\toprule
& {\bfseries$E_{01}$} & {\bfseries$E_{02}$} & {\bfseries $E_{10}$} & {\bfseries$E_{12}$} & {\bfseries$E_{20}$} & {\bfseries$E_{21}$} & {\bfseries$E_{\textnormal{Gross}}$} & {\bfseries$E_{\textnormal{Total}}$} & {\bfseries$E_{\textnormal{fine}}$} \\
\midrule
SD-fHMM & 3.26\% & 1.28\% & 0.28\% & 12.13\% & 0.22\% & 1.40\% & 0.02\% & 18.59\% & 1.75\%\\

SI-fHMM & 2.71\% & 1.18\% & 0.48\% & 11.20\% & 0.23\% & 1.85\% & 0.02\% & 17.67\% & 1.84\%\\

online-AMTC & 1.42\% & 0.30\% & 2.85\% & 1.73\% & 0.36\% & 8.00\% & 0.02\% & 14.64\% & 1.67\%\\

offline-AMTC & 1.49\% & 0.32\% & 2.71\% & 2.26\% & 0.41\% & 7.18\% & 0.03\% & 14.40\% & 1.80\% \\
\bottomrule
\end{tabular}
\end{table*}

\begin{table}
\centering
\ra{1.3}
\caption{Average computation time in seconds per $100$ frames}
\label{table_AvsF_CompTime}
\centering
\begin{tabular}{@{}rrr@{}}\toprule
& {\bfseries \textit{mixmax} likelihood (s)} & {\bfseries Tracking (s)}\\
\midrule
SD-fHMM & 39.47 & 3.96\\
SI-fHMM & 195.86 & 4.30\\
offline-AMTC & N/A & 0.10\\
online-AMTC & N/A & 0.44\\
\bottomrule
\end{tabular}
\end{table}

\subsubsection{Multiple Traces}\label{sssec::multitraceExp}
In this section, we evaluate the performance of the offline- and online-AMTC using simulated data in the presence of multiple traces and compare them with the fHMM method. To allow a fair comparison of our methods with fHMM, we adopt the performance measure proposed in~\cite{wu2003multipitch} with a slight change. We give details on our experiment setup as well as the error measure below.
\par To test both algorithms, we generated a corrupted frequency signal $s[n]$ with two frequency traces, \textit{i.e.,} $s[n] = \sum_{l=1}^2\sin \Phi_{(l)}[n]+\epsilon[n]$. The model trained in still mode for generating $\Phi[n]$ in Section~\ref{sssec::single_trace} is adopted for synthesizing both traces. The variance of $\epsilon[n]$ is tuned to achieve six SNR levels from $0$ to $-10$~dB. To cope with the high computational cost associated with running fHMM at a full scale, we cut signals to 1 minute, set the number of frequency bins to $64$, and made neighboring frequency bins $1$ bpm apart. The cardinality of frequency state was set to $169$ so that it uniformly covers the whole frequency range of interest. For each trace, we also introduced a $20$ seconds unvoiced segment. The starting point of the segment is drawn uniformly from the interval $[20, 30]$ s in the one-minute long signal.
\par We estimate the GMM parameters of the single-trace observation probabilistic model in the fHMM framework using the EM algorithm \cite{dempster1977maximum}. At each SNR level, we generated 6000 spectrum frames with a single frequency component for each $169$ frequency states (where the first state encodes unvoiced decision). We set the maximum number of components per GMM to 20 and used MDL \cite{wohlmayr2011probabilistic} to determine the number of components automatically. The parameters were trained in an SNR-dependent (SD) and an SNR-independent (SI) fashion (\textit{i.e.}, each SD model was trained only with samples of the corresponding SNR, and the SI model was trained with all samples). We adopted the mixture-maximization interaction model proposed in \cite{wohlmayr2011probabilistic}, and set the prior distribution for both fHMM and AMTC uniformly as $P(f_{(l)}(1)=m)=1/169$, $\forall m$, and the transition probability follows a uniform distribution with width parameter $k=2$. Moreover, the voiced to unvoiced transition probability for fHMM was empirically selected as $P(\text{voiced}|\text{unvoiced})=0.2$, and $P(\text{unvoiced}|\text{voiced})= 0.1$. 
\par To compare the tracking performance, we use the well-adopted error measure proposed in \cite{wu2003multipitch} as described below: 
\begin{itemize}
\item $E_{ij}$: the percentage of time frames where $i$ frequency components are misclassified as $j$.
\item $E_{\textnormal{\textnormal{Gross}}}$: the percentage of frames where $\exists l$, s.t. $\Delta f_{(l)}> 20\%$. We define the relative frequency deviation $\Delta f_{(l)}\triangleq\underset{i}{\min}\frac{|\hat{f}_i-f_{(l)}|}{f_{(l)}}$, and $f_{(l)}$ is the reference frequency for $l$th component.
\item $E_{\textnormal{fine}}^l$: the average relative frequency deviation from the reference of the $l$th frequency component for those frames where $\forall l$, $\Delta f_{(l)} \leq 20\%$. 
\end{itemize}
Note that both $E_{ij}$ and $E_{\textnormal{Gross}}$ represent a frame counting measure. We therefore group them together to form the total gross error: $E_{\textnormal{Total}} = E_{01}+ E_{02}+ E_{10}+ E_{12}+ E_{20}+ E_{21}+E_{\textnormal{Gross}}$, and define $E_{\textnormal{fine}} = E_{\textnormal{fine}}^1+E_{\textnormal{fine}}^2$.

\begin{figure}
\centering
\subfigure{\includegraphics[width=3.5in]{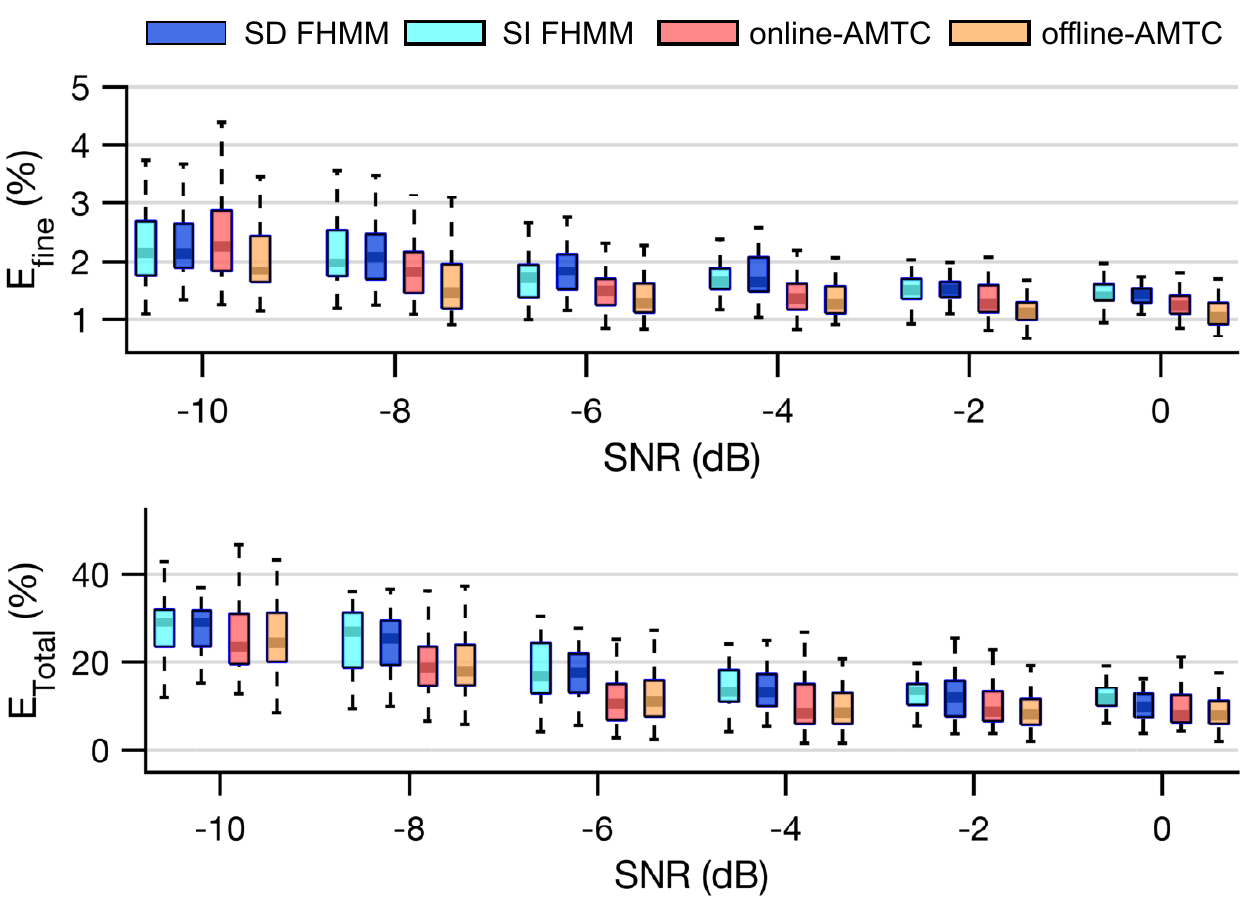}%
\label{subfig_MultiTrace_box_Etotal}}
\caption{Box plots of $E_{\textnormal{fine}}$ (top) and $E_{\textnormal{Total}}$ (bottom) of the two-trace tracking performance of} SD-fHMM, SI-fHMM, online-AMTC, and offline-AMTC at different levels of SNR.\label{fig_multitrace_box}
\end{figure}
\begin{figure}
\centering
\includegraphics[width=3.5in]{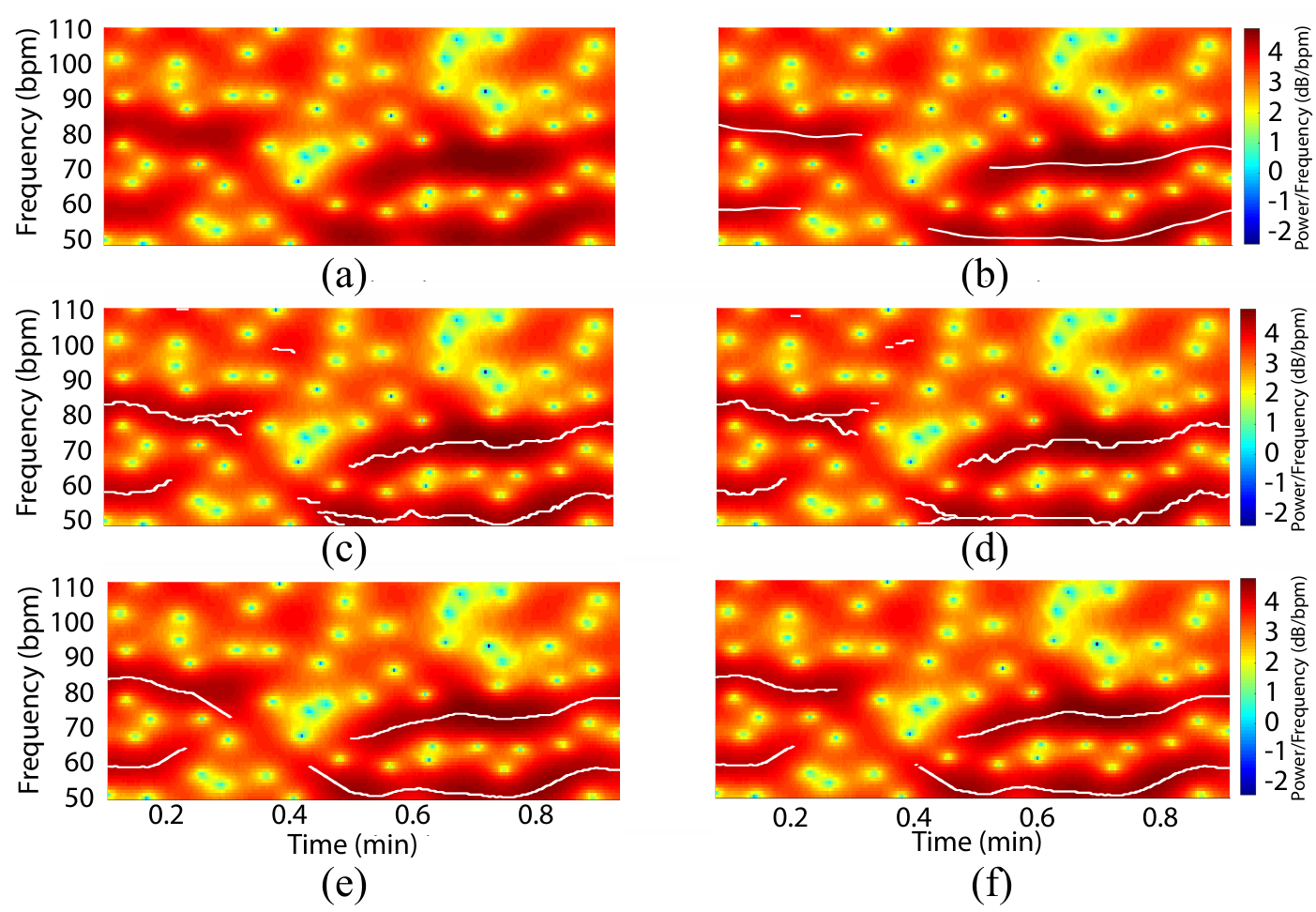}%
\label{fig_multi_ori}
\caption{(a) Spectrogram of one test instance with SNR $=-8$~dB; (b) same spectrogram overlaid by ground-truth traces. Tracking results by (c) SD-fHMM, (d) SI-fHMM, (e) offline-AMTC, and (f) online-AMTC.}
\label{fig_multi_example}
\end{figure}
\begin{figure}
\centering
\subfigure[]{\includegraphics[width=1.7in]{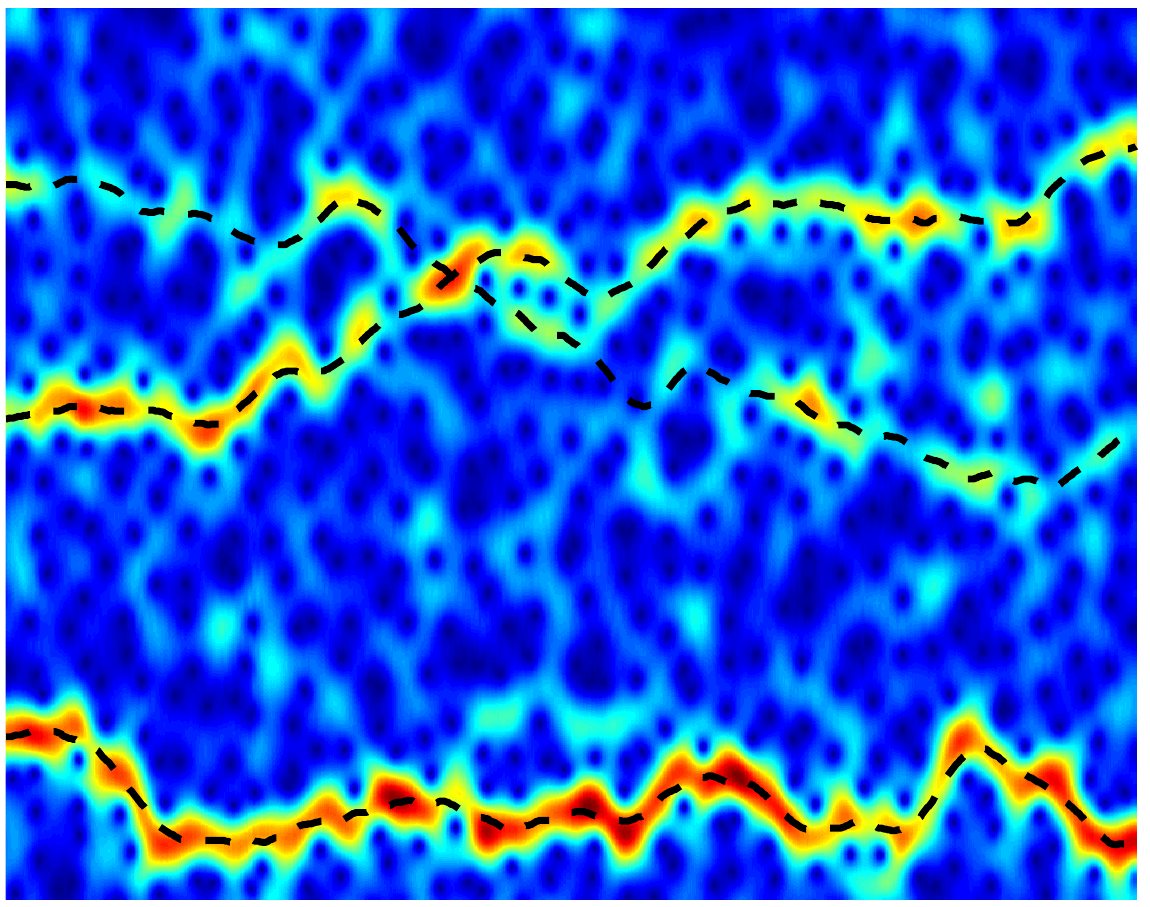}%
\label{subfig_MutiTrace_GT_03}}
\subfigure[]{\includegraphics[width=1.7in]{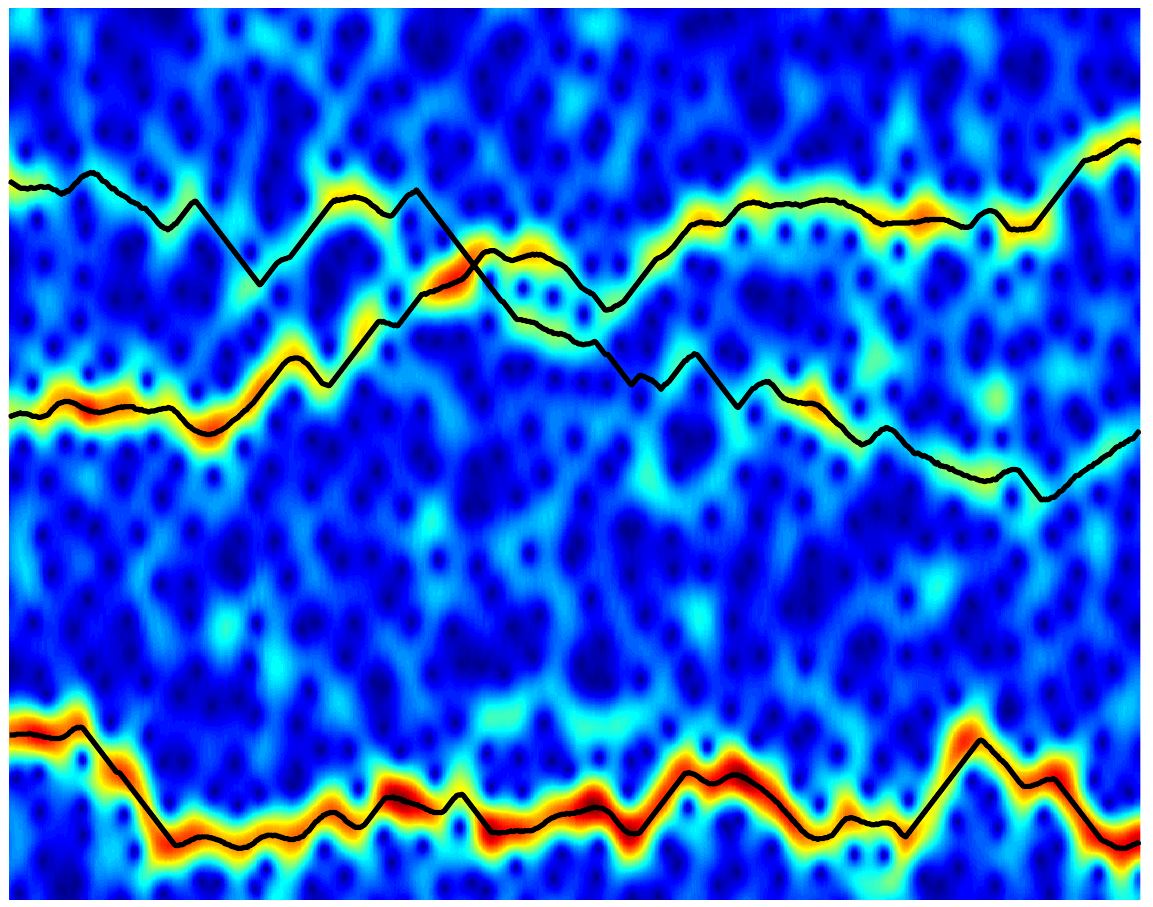}%
\label{subfig_MultiTrace_AMTC}}
\caption{(a) Ground-truth frequency traces at $-10$~dB in the spectrogram of a synthetic signal. (b) Three estimated traces by AMTC.}\label{fig_multitrace_example}
\end{figure}
\begin{figure}
\centering
\subfigure[]{\includegraphics[width=1.7in]{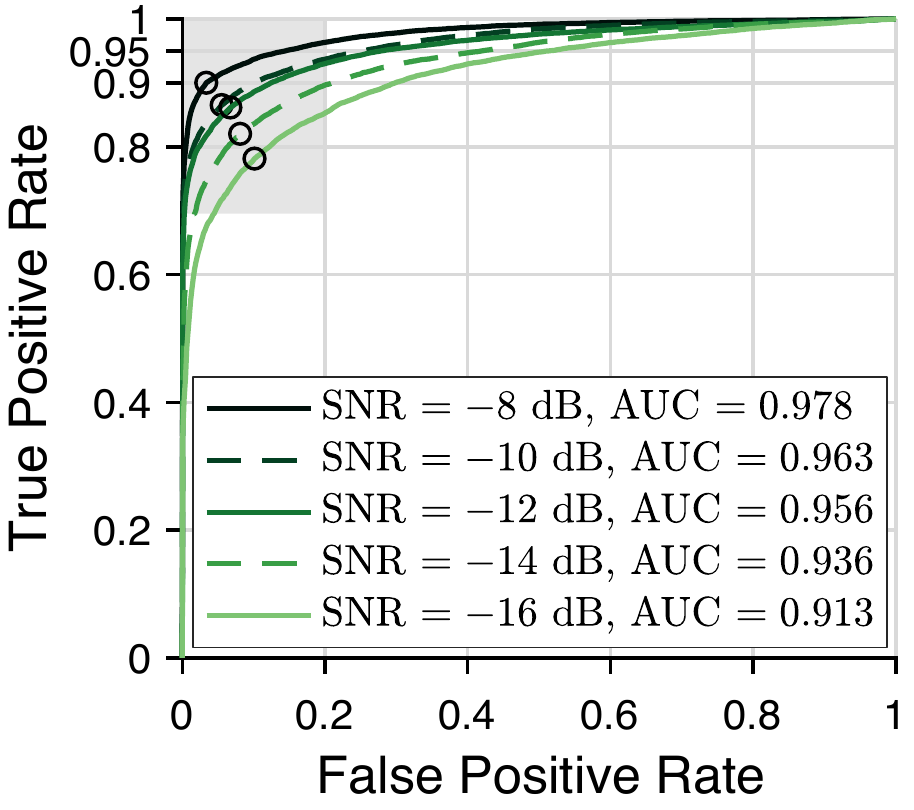}}%
\subfigure[]{\includegraphics[width=1.7in]{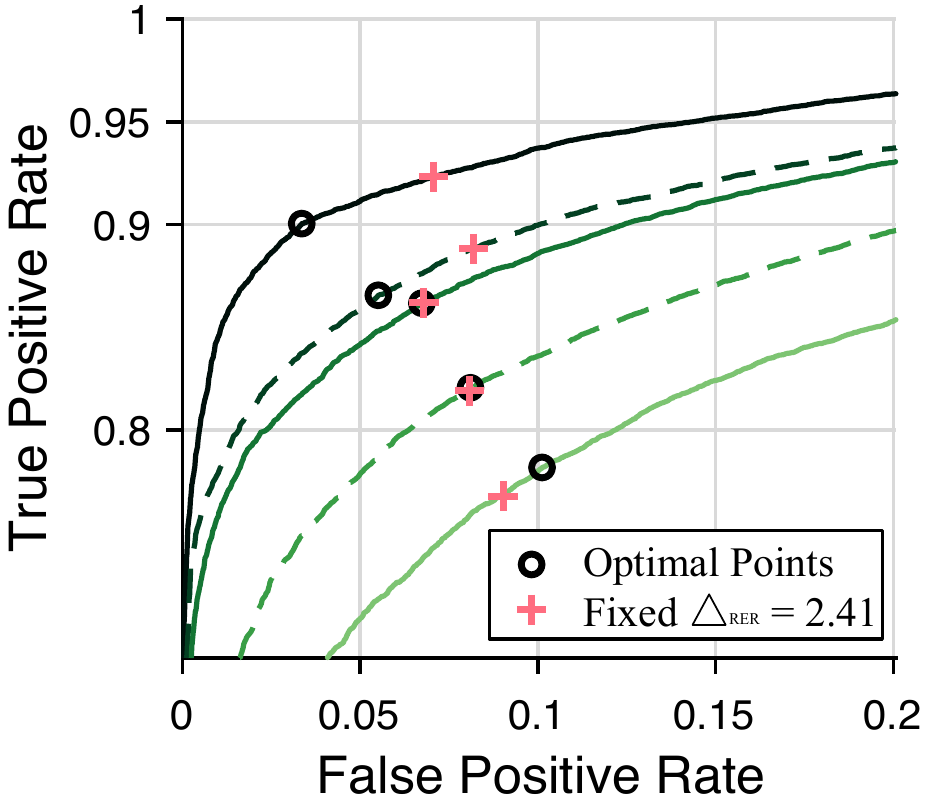}}%
\caption{(a) ROC curves for the proposed trace detection method at different SNR levels. (b) The zoomed-in plot of the shaded area in (a) with optimal operating points (black circle) and operating points using fixed threshold ($\Delta_{\text{RER}}= 2.41$, pink plus sign).}\label{fig::detect_ROC}
\end{figure}

\par To test the performance, we generated $300$ test signals for each SNR level using the same setting mentioned above. We compared the performance of SD-fHMM, SI-fHMM, offline-AMTC and online-AMTC using the aforementioned error measures and the results are listed in Table~\ref{table_multi_trace_err}. We depict the distribution of $E_{\textnormal{Total}}$ and $E_{\textnormal{fine}}$ specifically in Fig.~\ref{fig_multitrace_box}. All methods have a similar performance in terms of the fine detection error $E_{\textnormal{fine}}$, while AMTC slightly outperforms fHMM in terms of $E_{\textnormal{Total}}$, the main contributor of which is $E_{12}$. Table~\ref{table_AvsF_CompTime} shows the average computation time for the \textit{mixmax} likelihood estimation procedure~\cite{wohlmayr2011probabilistic}, together with the tracking time requirement tested on a 2014 MacBook Pro with a $2.3$~GHz Intel Core i5 processor. Note that the preprocessing stage of fHMM to compute the emission probability also consumes almost $0.4$ s/frame for the SD and $2.0$ s/frame for the SI model, which makes the real-time implementation almost impossible for a usual hardware setting. AMTC, on the other hand, is much more computationally efficient than fHMM even without considering the \textit{mixmax} likelihood computing. For this task, the online-AMTC reported a similar performance compared with the offline version at $4.4$ ms/frame. It guarantees real-time adaptation with almost no performance drop. Fig.~\ref{fig_multi_example} shows the experimental results of the proposed algorithm and fHMM on a test signal with  $\textrm{SNR}= -8$~dB. We can observe that in a low SNR environment, the performance of the offline- and online-AMTC are better than fHMM algorithm in terms of accuracy and false-positive detections.
\par Fig.~\ref{fig_multitrace_example}(b) shows an example of the tracking result of the offline-AMTC when SNR is $-10$~dB and three traces are presented. We can see three traces have been accurately estimated as compared to the ground truth on the left when two weak traces with different levels of strength intersect.

\subsubsection{Trace Detection}\label{sssec::trace_detection}
In this part, we evaluate the trace detection performance and the optimal selection of threshold $\Delta_{\text{RER}}$ using the synthetic data under five SNR levels. We generated $100$ trials for each level of SNR with the generative model described in Section~\ref{sssec::single_trace}. An unvoiced segment was inserted in each test signal with the starting \rev{time position} randomly selected from the signal. The length of the selected segment ranged from $25\%$ to $75\%$ of the signal length, and, over the whole dataset, the number of voiced spectral frames equaled the number of unvoiced frames. In this experiment, the voiced detection is treated as the positive case, and the detection result (without the postprocessing operation using $\Delta_1$ and $\Delta_2$) is summarized using the Receiver Operating Characteristic (ROC) plot in Fig.~\ref{fig::detect_ROC}(a). From the plot, we observe highly accurate detection results for each SNR condition with the Area Under the Curve (AUC) higher than $0.9$. 

In Fig.~\ref{fig::detect_ROC}(b), we show the zoomed-in plot of the shaded area in Fig.~\ref{fig::detect_ROC}(a). The optimal operating points in terms of minimizing the sum of false negative and false positive rate are shown in black circles. The operating point corresponding to a fixed threshold, namely, $\Delta_{\text{RER}}=2.41$ (the value we used for the experiments in the paper), are also shown using pink plus signs. Note that the detection results using a fixed threshold value are close to those with the optimal choice at every SNR level, demonstrating that the chosen value of $\Delta_{\text{RER}}$ is effective and almost independent of the SNR.

\subsection{Experimental Results on rPPG Data}
We evaluated the performance of the proposed method on a real-world dataset from the problem of the pulse rate estimation from facial videos. We show by experiment that AMTC can successfully extract the subtle pulse trace even when the trace is dominated by another frequency component. To test the robustness of the algorithm in a challenging situation, we use the dataset where the video contains significant subject motion~\cite{zhu2017ICIP}. In total, the dataset contains 20 videos in which 10 contain human motions on an elliptical machine, and the other 10 contain motions on a treadmill. Each video is about 3 minutes long in order to cover various stages of fitness exercise. Each video was captured in front of the subject's face by a commodity mobile camera (iPhone 6s) affixed on a tripod or held by the hands of a person other than the subject. The heart rate of the test subject was simultaneously monitored by an electrocardiogram (ECG)-based chest belt (Polar H7) for reference. The spectrogram of the preprocessed face color feature was estimated using the same set of parameters as in Section~\ref{sssec::single_trace}. The estimated SNR of the dataset is $-6.31$~dB using the estimation method introduced in~\cite{de2013RobustRPPGusingChrom}.

\begin{figure}
\centering
\includegraphics[width=3.5in]{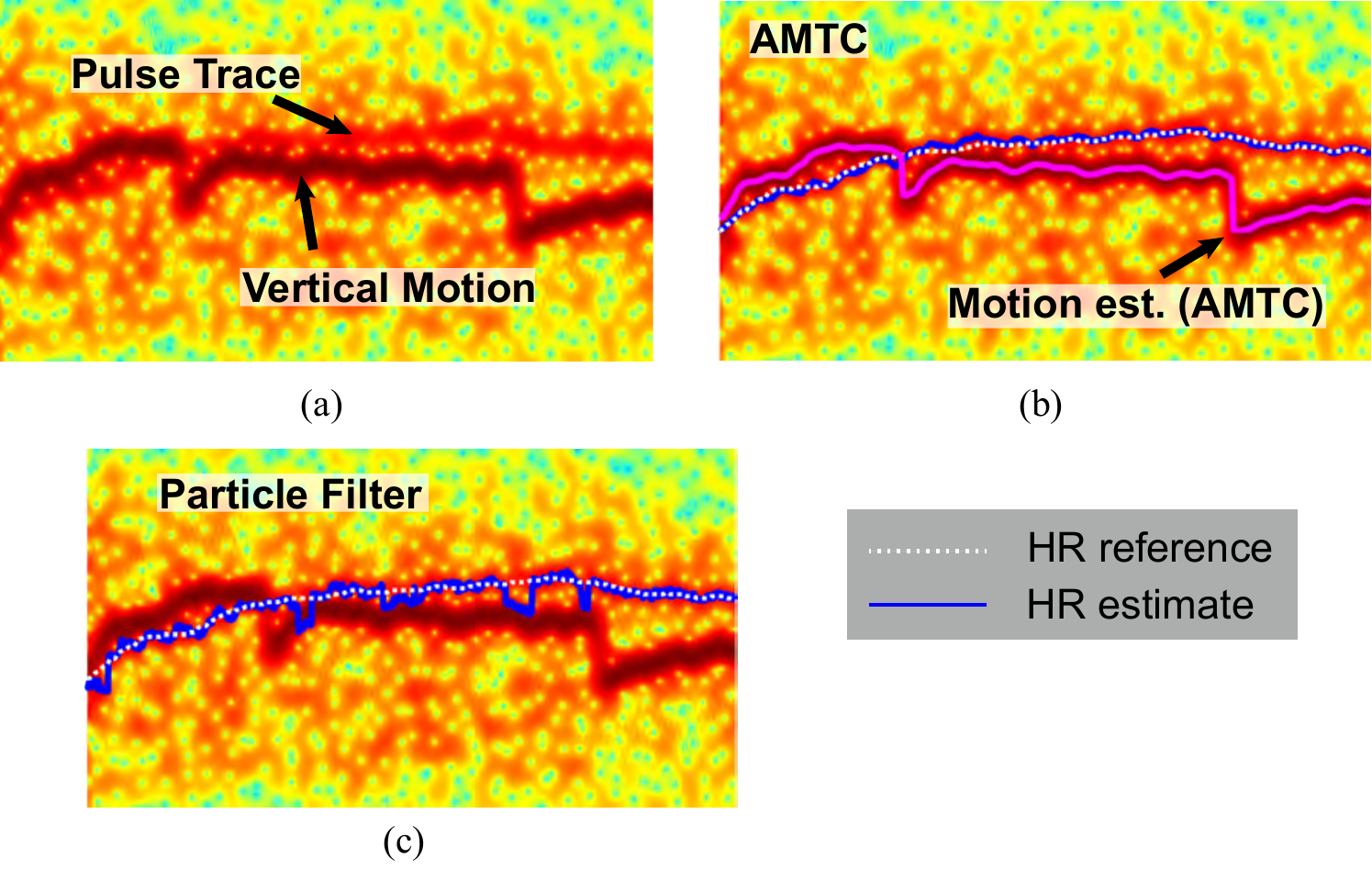}%
\caption{(a) A weak heart-rate trace dominated by a strong trace induced by vertical motion of the person running on an elliptical machine. The estimated pulse SNR equaled $-4.5$~dB. (b) Heart rate estimation after compensating the first trace estimate using the offline-AMTC. (c) Heart rate estimation using motion spectrogram notching and particle filter method. The estimation result is compared with the heart rate (white dashed line) simultaneously measured by an electrocardiogram-based sensor.}\label{fig_sim_example}
\end{figure}

Fig.~\ref{fig_sim_example} gives an example of the tracking result using AMTC with a uniform Markov transition probability model with $k=60$ for first motion-induced trace estimate and with $k=2$ for second pulse-induced trace estimate. More freedom of trace dynamic ($k=60$) was assigned to the first estimate as the variation of motion frequency can be much greater than the heart rate. We notice that for each spectrogram, the traces induced by subject motions dominate the heart rate trace. Compared to the particle filter-based method that utilizes additional information to compensate for the motion trace~\cite{zhu2017ICIP}, AMTC can faithfully track the dominating motion trace and recognize the PR trace as the second trace. Notice that the trace estimate from the particle filter would occasionally deviate to the vertical motion trace. We summarize the sample mean $\hat{\mu}$ and standard deviation $\hat{\sigma}$ of the error measures for all of our videos, and the results are listed in Table~\ref{tab::rPPG}. The average error for AMTC is $2.21$~bpm in the offline mode and $2.78$~bpm in the online mode in \textsc{Rmse}, and $3.16\%$ in the offline mode and $4.01\%$ in the online mode in relative error. The performance of AMTC is more than twice as high compared to that of the state-of-the-art approach employing motion notching and particle filter.

\begin{figure}
\centering
\includegraphics[width=3.5in]{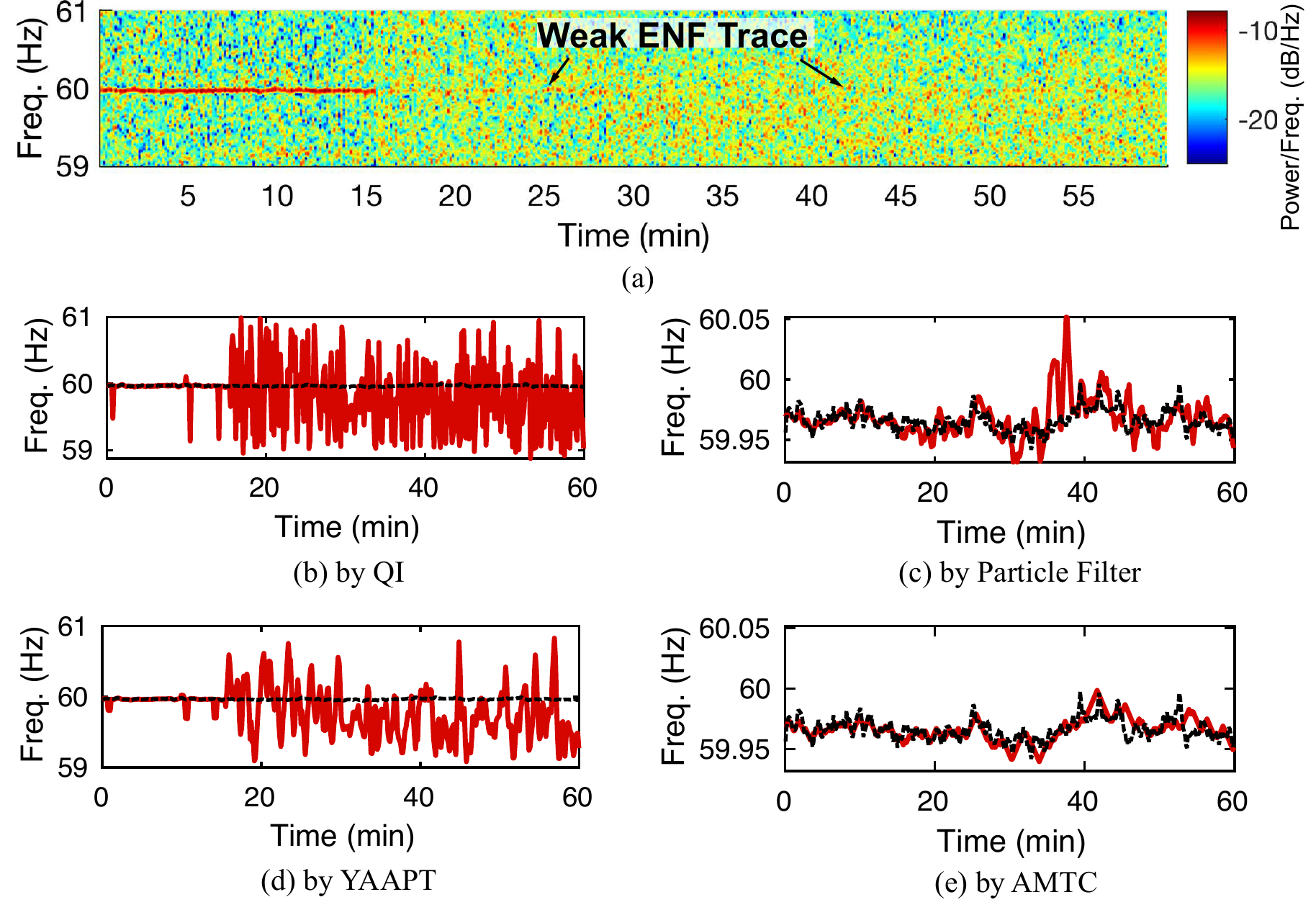}%
\caption{(a) Spectrogram for a sample ENF audio signal. The estimated ENF SNR is $-8.2$~dB. Trace estimates (red line) returned by (b) Quadratic Interpolation, (c) Particle Filter, (d) YAAPT, and (e) offline-AMTC. The reference ENF trace is shown in dashed black line in plots (b)--(e).}\label{fig_ENFExample}
\end{figure}

\begin{table}[!t]
\centering
\ra{1.3}
\caption{Performance comparison of the proposed method and the particle filter method on rPPG data}
\label{tab::rPPG}
\begin{tabular}{@{}rrrrrrr@{}}\toprule
& \multicolumn{2}{c}{\textsc{Rmse} (bpm)} & \multicolumn{2}{c}{\textsc{ERate} (\%)} & \multicolumn{2}{c}{\textsc{ECount} (\%)}\\ \cmidrule(lr){2-3} \cmidrule(lr){4-5} \cmidrule(lr){6-7}& \multicolumn{1}{c}{$\hat{\mu}$} & \multicolumn{1}{c}{$\hat{\sigma}$} & \multicolumn{1}{c}{$\hat{\mu}$} & \multicolumn{1}{c}{$\hat{\sigma}$} & \multicolumn{1}{c}{$\hat{\mu}$} & \multicolumn{1}{c}{$\hat{\sigma}$}\\ \midrule
MN+PF & 5.29& 5.51& 9.41 & 14.13 & 2.20 & 2.24\\
offline-AMTC & 2.21& 1.11& 3.16 & 6.04 & 1.02 & 2.24\\
online-AMTC & 2.78& 1.20 & 4.01 & 6.42 & 1.25 & 2.42\\\bottomrule
\end{tabular}
\end{table}

\begin{table}[!t]
\centering
\ra{1.3}
\caption{Performance of various methods on ENF data}
\label{tab::ENF}
\begin{tabular}{@{}rrrrr@{}}\toprule
& \multicolumn{2}{c}{\textsc{Rmse} (Hz)} & \multicolumn{2}{c}{Pearson's $\rho$} \\ \cmidrule(lr){2-3} \cmidrule(lr){4-5}
& \multicolumn{1}{c}{$\hat{\mu}$} & \multicolumn{1}{c}{$\hat{\sigma}$} & \multicolumn{1}{c}{$\hat{\mu}$} & \multicolumn{1}{c}{$\hat{\sigma}$}\\ \midrule
QI & 0.24 & 0.18 & 0.18 & 0.26\\
Particle Filter & 0.04 & 0.07 & 0.55 & 0.37\\
YAAPT & 0.16 & 0.12 & 0.23 & 0.28\\
offline-AMTC & 0.01 & 0.01 & 0.85 & 0.18\\
online-AMTC & 0.03 & 0.02 & 0.81 & 0.20\\\bottomrule
\end{tabular}
\end{table}

\subsection{Experimental Results on ENF Data}
In this subsection, we test the performance of the proposed algorithm on a real-world ENF dataset. In total, 27 pairs of one-hour power grid signal and audio signal from a variety of locations in North America were collected and tested. Each pair of signals were simultaneously recorded using a battery-powered Olympus Voice Recorder WS-700M at a sampling rate of $44.1$~kHz in MP3 format at $256$~kbps. All the audio signals were recorded in typical apartment rooms in U.S. with only ambient noise and ENF interference induced by the mains-powered appliances. We recorded the reference ENF signal from the power mains of the electrical supply. To limit the voltage to the safe range of the input of a sound card or a digital recorder, we used a step-down transformer to convert the power supply voltage level to $5$~V and then used a voltage divider with resistors of $33$~Ohm and $33$~kOhm to obtain an input of $5$~mV~\cite{garg2013seeing}.

Taking the collected audio recordings, we downsample the signals to $1$~kHz to reduce the computational load, and apply the \textit{harmonic combining method}~\cite{hajj2013spectrum} to obtain robust frequency strips around the nominal frequency, \textit{i.e.}, 60~Hz in North America. The STFT is performed with a rectangular window of $8$ seconds long, no overlap between adjacent frames, and neighboring frequency bins of $0.004$~Hz apart. The harmonic combining method exploits different ENF components appearing in a signal, and adaptively combines them based on the local SNR to achieve a more robust and accurate estimate than that by using only one component. We obtain the ground truth from the corresponding power grid signals using \textit{Quadratic Interpolation} (QI)~\cite{smith1987parshl}, as the SNR is high and frame-wise highest peak method is proved to be the maximum likelihood estimator of signal frequency~\cite{rife1974single}. According to the extracted ground-truth ENF, the averaged estimated SNR of the ENF signal is $-5.23$~dB using the estimation method introduced in~\cite{hajj2018factorAudioENFCapture}. We use \textsc{Rmse} and Pearson correlation coefficient $\rho$ of the estimated versus the ground-truth sequences of frequency variations as performance indices. They are two well-adopted error measures for ENF estimation. 
\par Fig.~\ref{fig_ENFExample} gives a tracking example using a piece of acoustic recording captured in San Diego, CA. Note that the ENF trace becomes weak after 15 minutes, which we define as a checkpoint. AMTC can identify the trace from the noisy harmonic combined spectrum feature. The particle filter approach gives comparable results before the checkpoint but deviates from the true trace occasionally due to nearby interference. Local peak based tracking method YAAPT and frame-wise frequency estimator QI completely lost the target after the checkpoint as the peak information alone is not able to guarantee a good estimate. 
\par The performance of various methods is summarized in Table~\ref{tab::ENF}. We calculate the sample mean and standard deviation of the error measures for 27 pieces of audio ENF signals. For this very noisy dataset, AMTC can achieve $0.01$~Hz in offline mode and $0.03$~Hz in online mode in average \textsc{Rmse} and $0.85$ in offline mode and $0.81$ in online mode in average correlation with ground truth, which outperforms all other tested tracking methods substantially both in average and variance of the error statistics. 

\begin{figure*}
  \centering

  \subfigure{\includegraphics[width=2.2in]{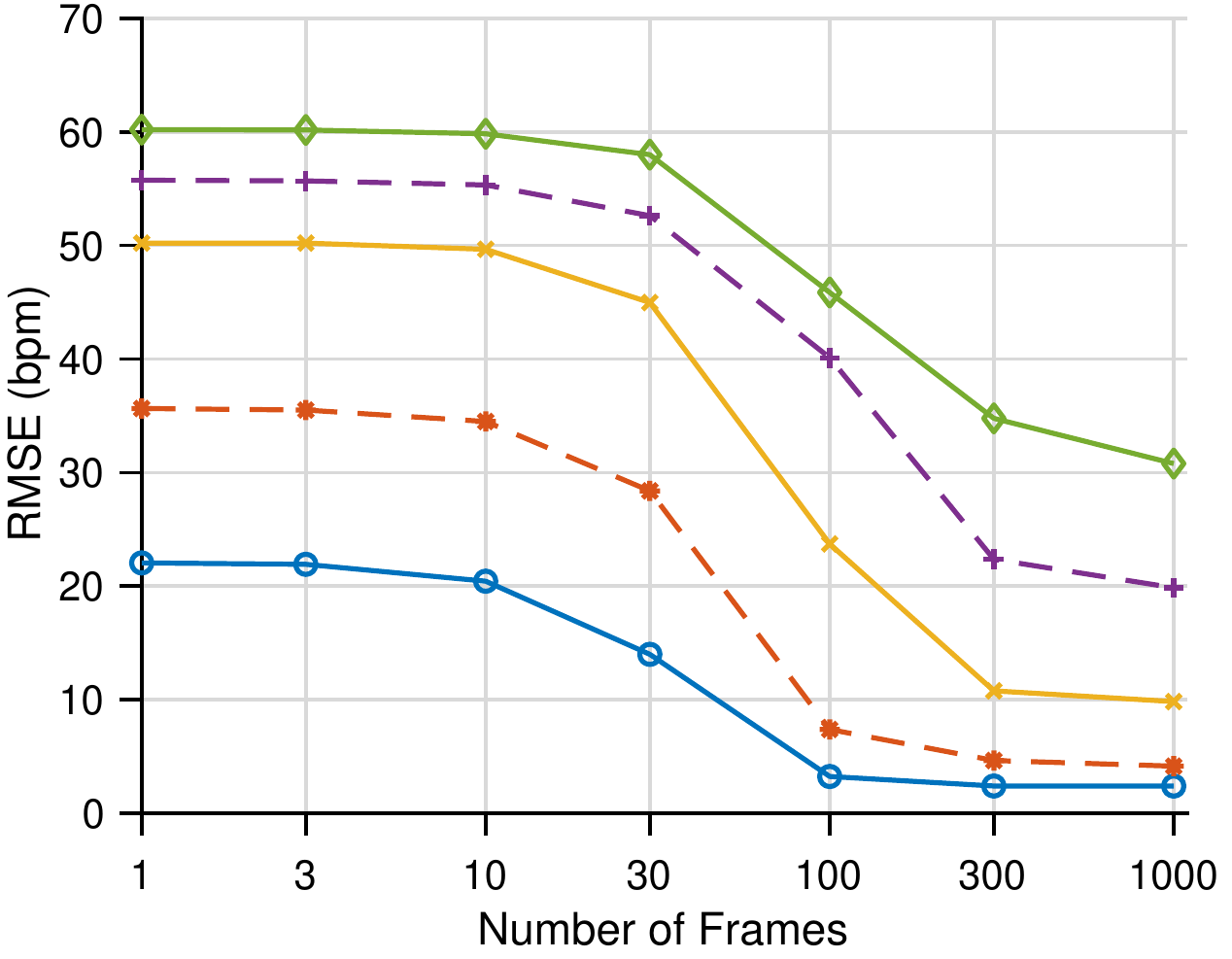}}%
  \subfigure{\includegraphics[width=2.2in]{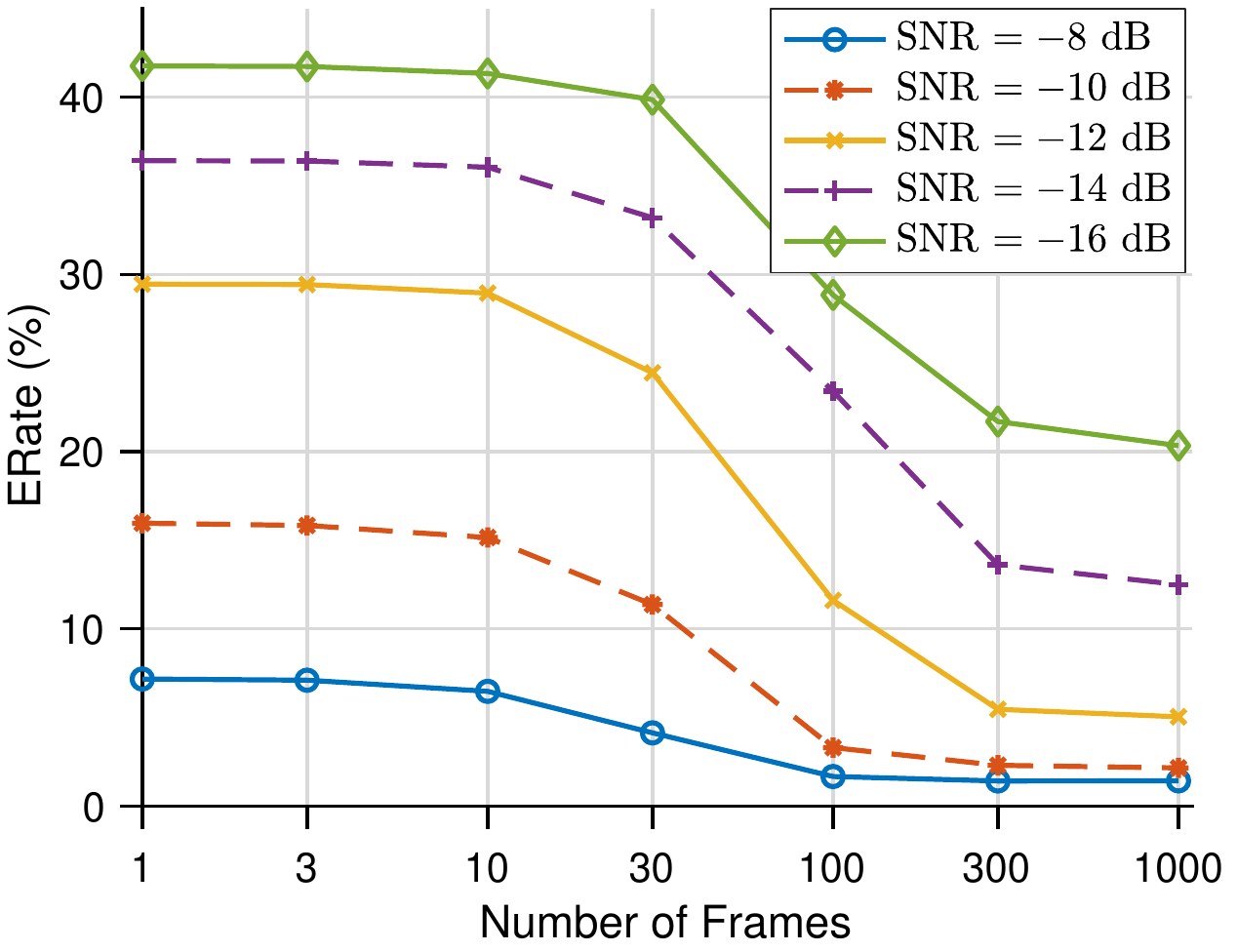}}%
  \subfigure{\includegraphics[width=2.2in]{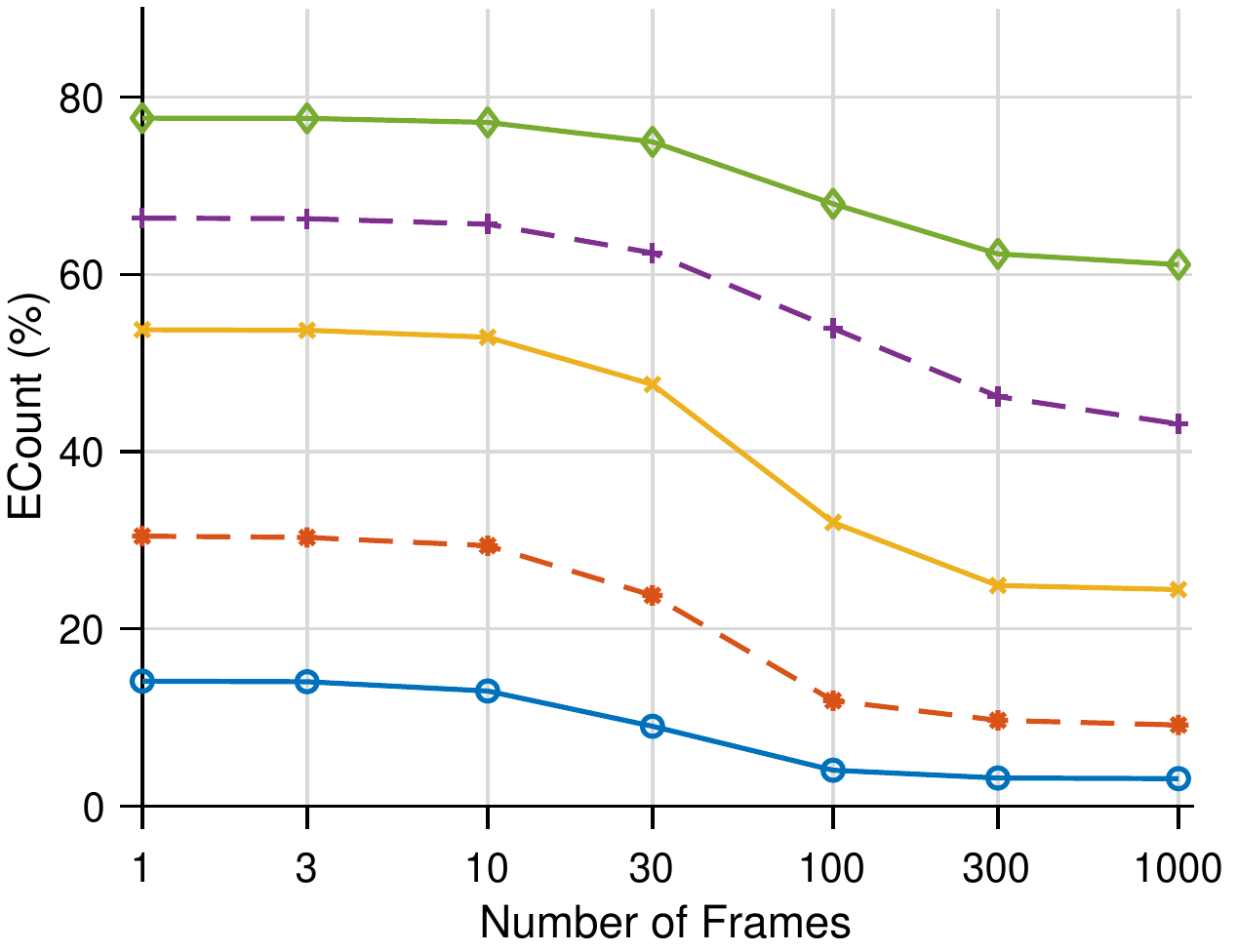}}\\%
  \subfigure{\includegraphics[width=2.2in]{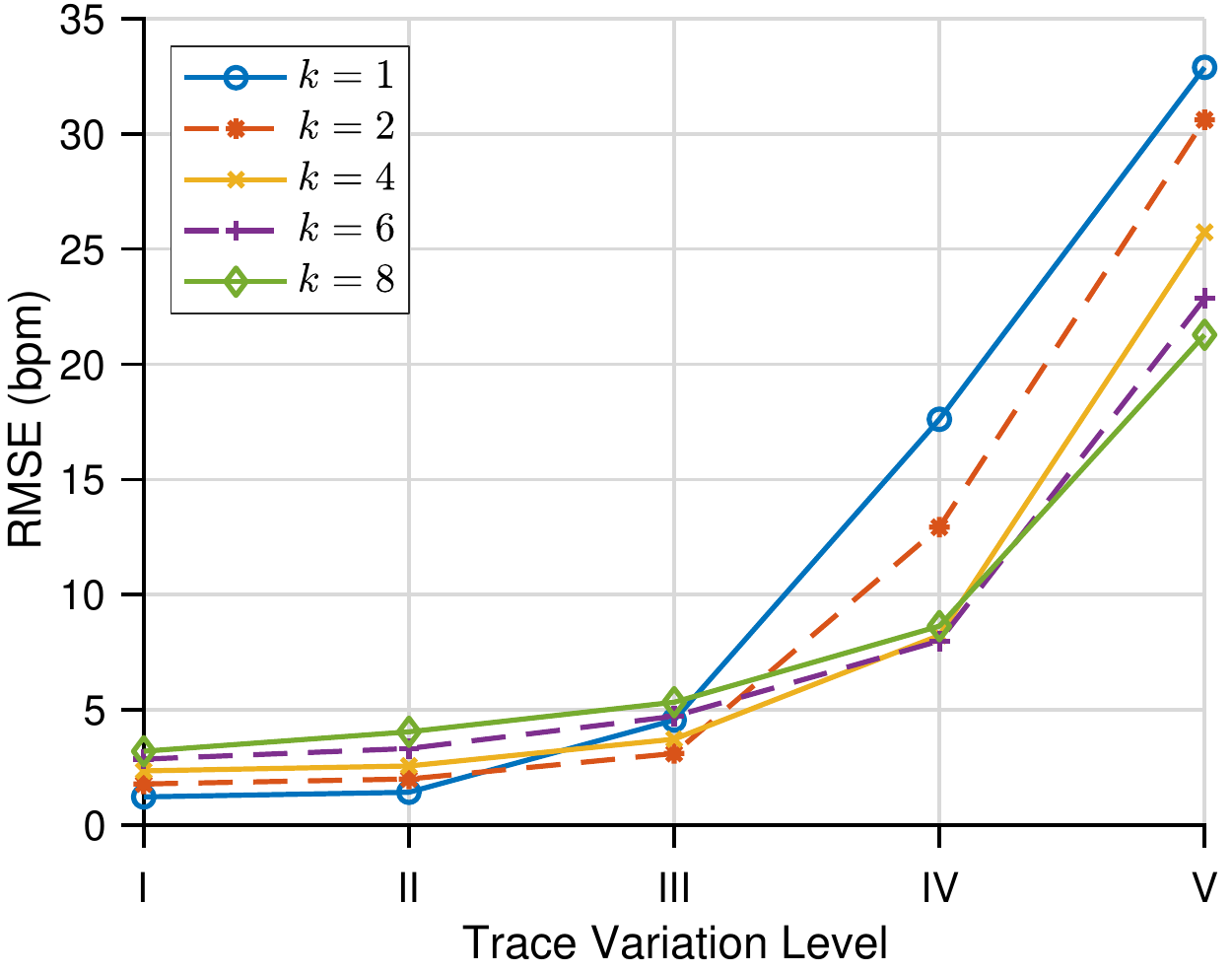}}%
  \subfigure{\includegraphics[width=2.2in]{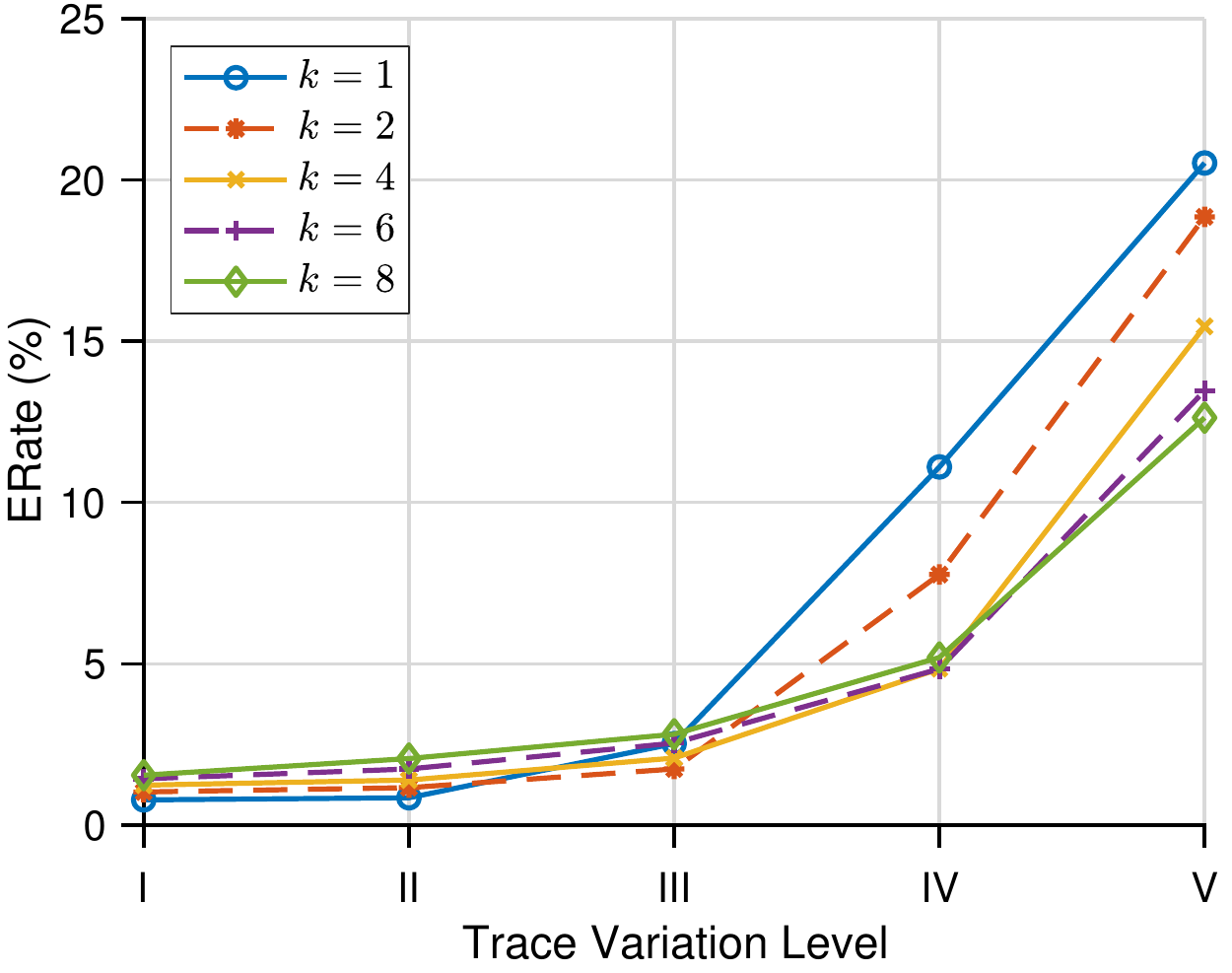}}%
  \subfigure{\includegraphics[width=2.2in]{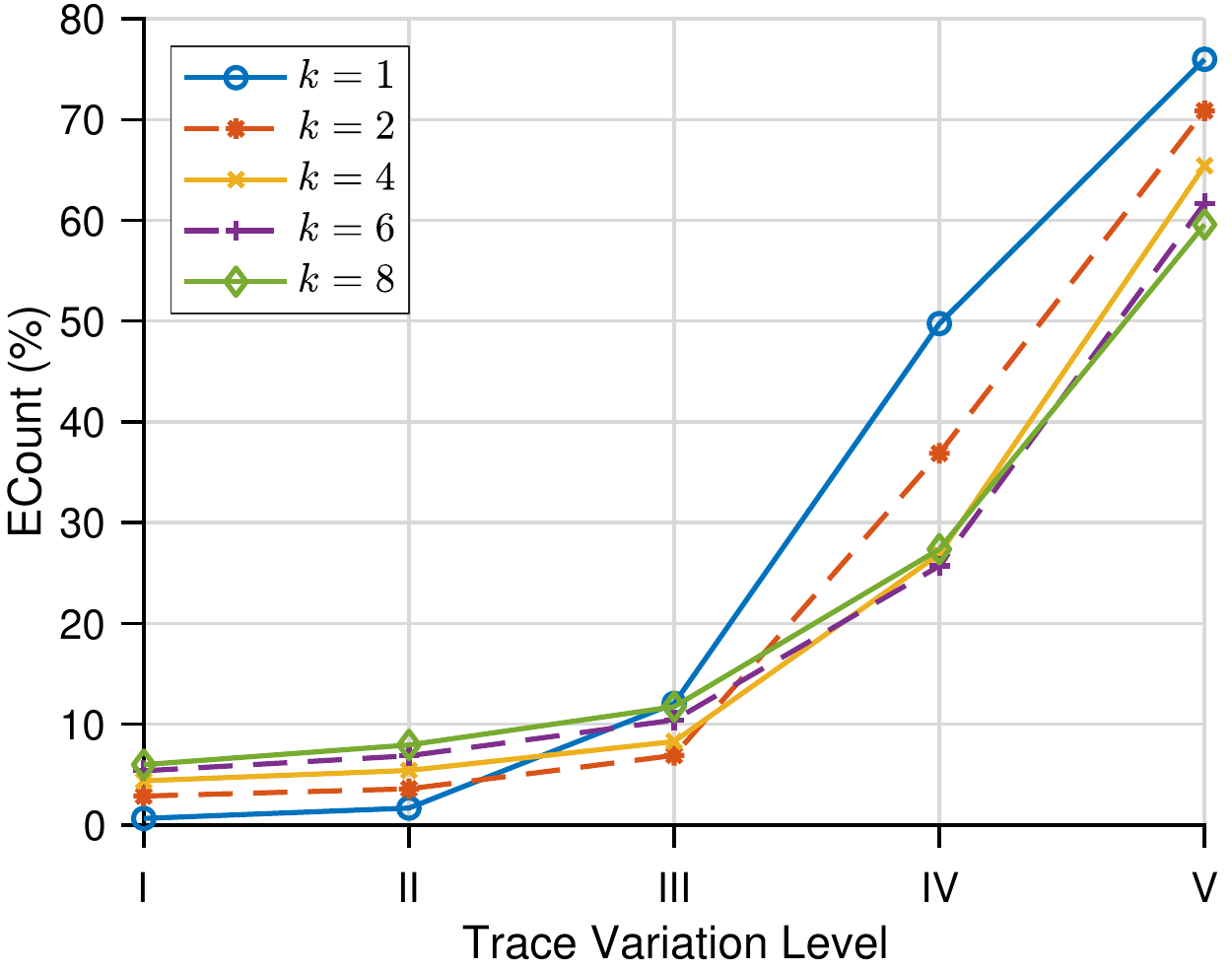}}\\%
  \subfigure{\includegraphics[width=2.2in]{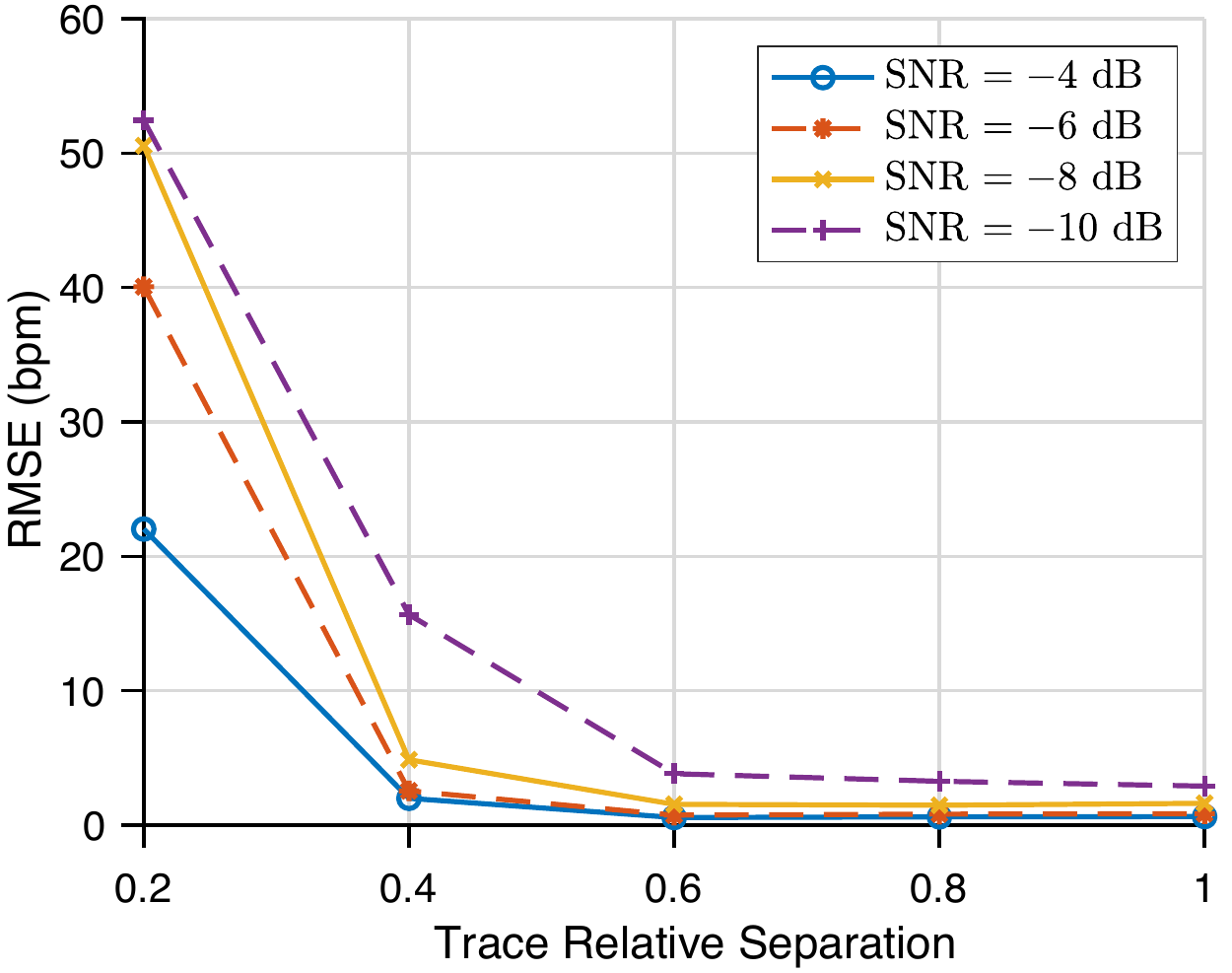}}%
  \subfigure{\includegraphics[width=2.2in]{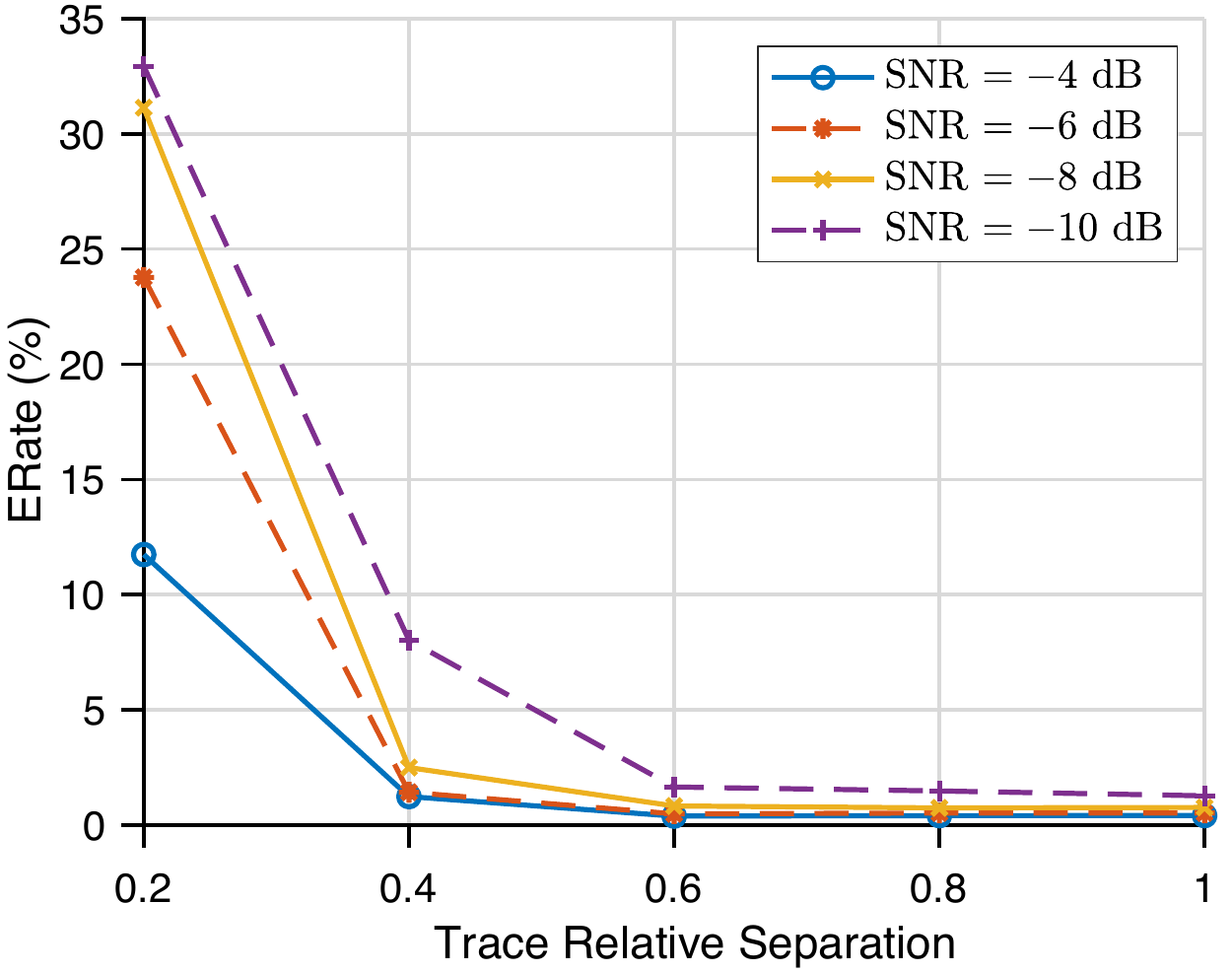}}%
  \subfigure{\includegraphics[width=2.2in]{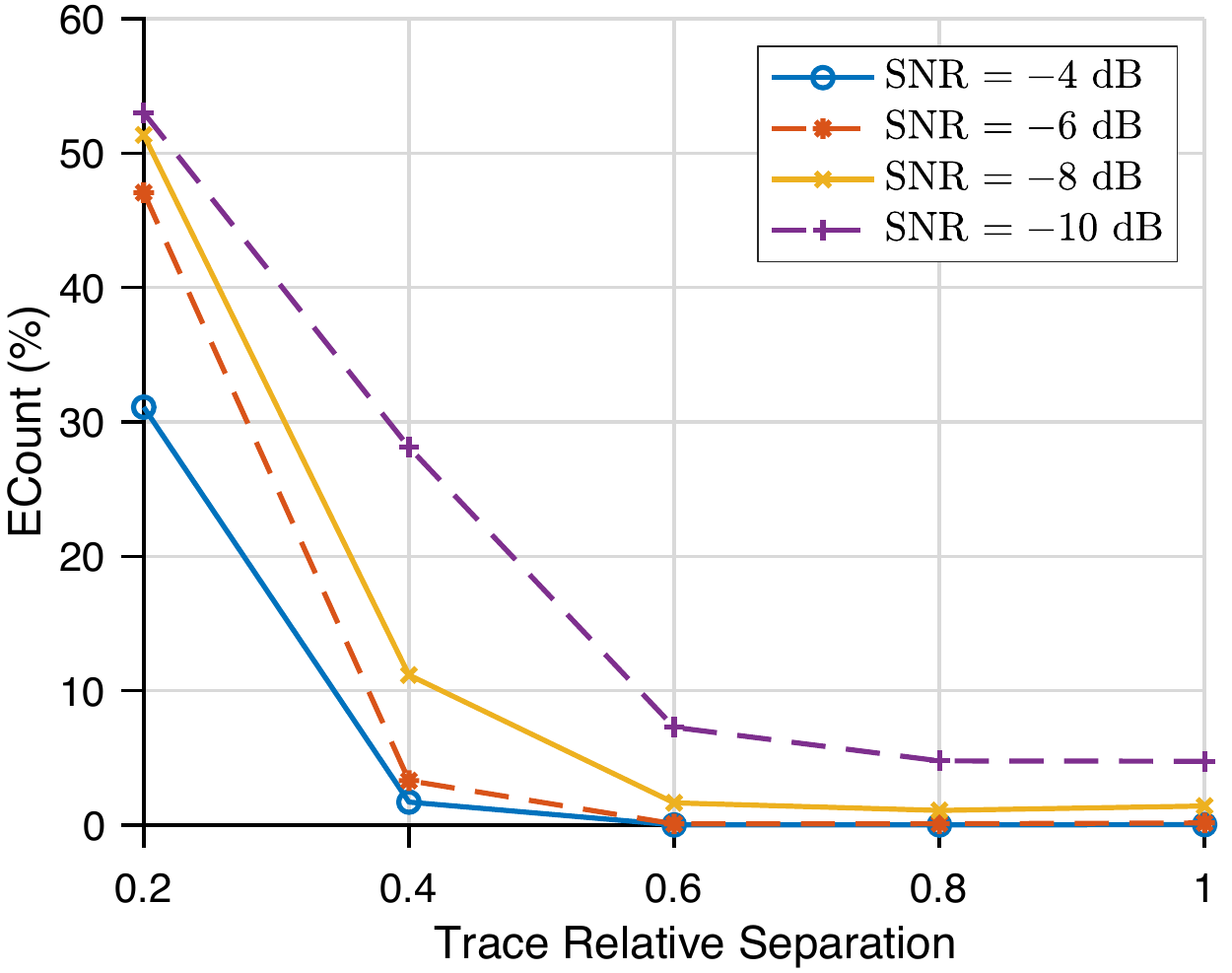}}%
  
  \caption{Impact evaluation of three factors: \textsc{Rmse} (first column), \textsc{ERate} (second column), and \textsc{ECount} (last column) of the trace estimates by the offline-AMTC as a function of signal length under different SNR scenarios (first row), as a function of different trace variation levels and selections of $k$ (second row), and as a function of TRS under different SNR scenarios (last row).}
  \label{fig::impact_factor}
\end{figure*}

\begin{figure}
\centering
\includegraphics[width=3.5in]{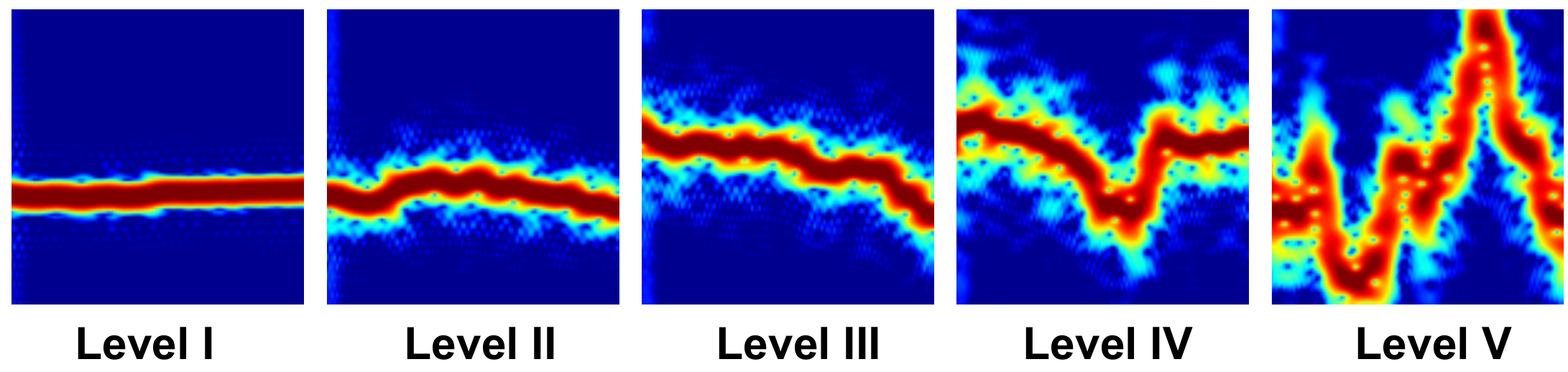}\\%
\caption{Spectrogram examples of raw signals with five trace variation levels before being further corrupted.}\label{fig::traceDynamic_example}
\end{figure}

\begin{figure}
\centering
\subfigure[]{\includegraphics[width=1.7in]{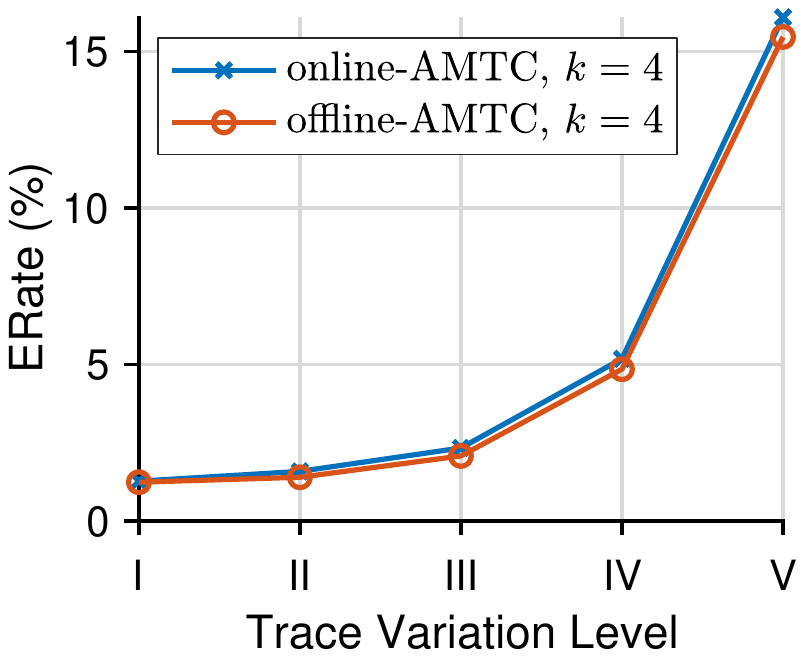}\label{subfig::impact_factor_onoff_transitionMismatch}}
\subfigure[]{\includegraphics[width=1.7in]{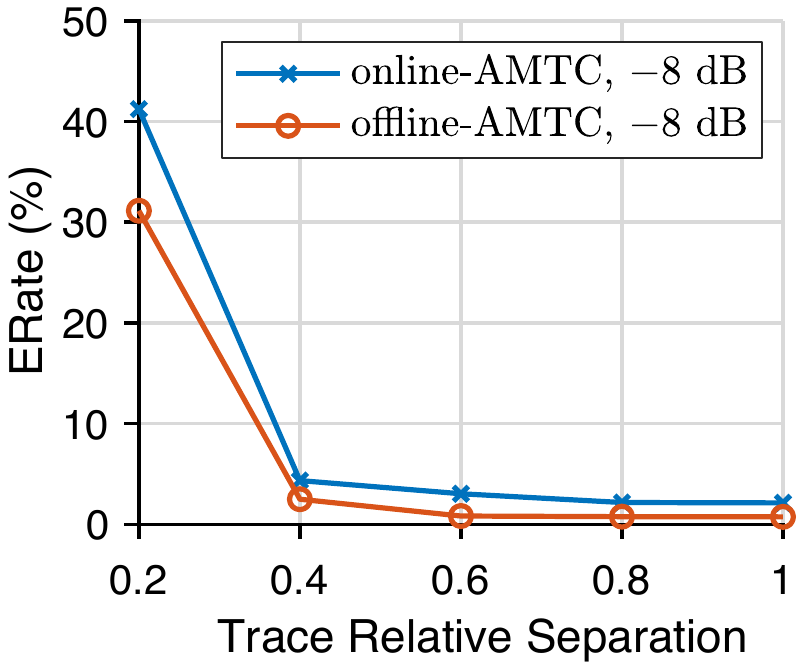}\label{subfig::impatc_factor_onoff_closeTrace}}
\caption{\textsc{ERate} of the trace estimates by the online-AMTC and the offline-AMTC (a) as a function of different trace variation levels when $k=4$, and (b) as a function of TRS when SNR$=-8$~dB.}
\label{fig::impact_factor_OnOff}
\end{figure}
\section{Impact of Various Factors}\label{sec::impact_factors}
In this section, we further evaluate the impact to the performance due to various factors. First, we study the performance when the number of spectral frames varies. Then, we discuss the effect of the trace variation level. Finally, we evaluate the impact of the separation between two traces to the estimation accuracy. The parameters are configured to be the same as introduced in Section~\ref{sssec::single_trace} unless otherwise stated.

\subsection{Impact of the Number of Frames}\label{ssec::impact_sig_len}
A frequency tracker starts to produce a meaningful tracking result by using two or more frames, and it is generally expected to have an improved tracking performance when more frames are used. This can be seen from the information theoretic viewpoint. Consider the true frequency state $f$ at the time instant $n$ as a random variable and denote noisy observed data at $n$th frame by $\mathbf{Z}(n)$. Using the ``conditioning reduces entropy'' lemma~\cite{cover-thomas2012info_theory} from information theory, we obtain the relationship between two posteriors $H(f(n)|\mathbf{Z}(n),...,\mathbf{Z}(2),\mathbf{Z}(1)) \leq H(f(n)|\mathbf{Z}(n),...,\mathbf{Z}(2))$, where $H(\cdot|\cdot)$ is the conditional entropy, suggesting less or equal uncertainty in $f(n)$ when more observations/frames are included during an inference process. Below we use experimental results to confirm that more accurate tracking results are achieved when the number of frames in the spectrogram increases.

We generate $200$ trials under SNR conditions at $-16$~dB, $-14$~dB, $-12$~dB, $-10$~dB, and $-8$~dB. The duration of the test signal is set to three and a half minutes, which is equivalent to $1000$ spectral frames in the spectrogram. The $1000$ spectral frames are then segmented uniformly in time without overlap based on the seven levels of evaluated number of frames, and the offline-AMTC is performed independently in each segment. Three performance metrics with respect to the number of frames under different SNR levels are shown in the first row of Fig.~\ref{fig::impact_factor}. Note that when the number of frame equals one, the tracking result using AMTC degenerates to the highest peak method. We can observe from the plots that the performance of the algorithm improves significantly when the signal length exceeds $10$~frames. More frames are needed in a lower SNR condition to reach a given performance level, whereas the performance starts to converge when the number of frames reach $300$ for all SNR levels.

\subsection{Impact of Trace Variation}\label{ssec::impact_trace_var}
During the formulation process of the frequency trace tracking problem, we have assumed the change of the frequency value between two consecutive bins as a one-step discrete-time Markov chain, characterized by a transitional probability matrix $\mathbf{P}$. With a training dataset of sufficient size available to the user, one may learn the model parameters of $\mathbf{P}$ to make a more precise tracking estimation. However, the training set is often unavailable in a real-world setting, and the user has to make their own choice of the $\mathbf{P}$ before deploying the algorithm. It is therefore important for a robust frequency tracker to successfully track the frequency components even when the variation of the frequency traces is at different levels.

We evaluate the system performance for both offline-AMTC and online-AMTC with respect to five different trace variation levels, and assume the transition probability follows the uniform distribution parameterized by $k$. $200$ trials are generated for each level of trace variation by tuning the variance of $f[n]$ in the generative signal model described in Section~\ref{sssec::single_trace}. Specifically, the five levels of the trace variation correspond to $0.001$, $0.005$, $0.01$, $0.02$, and $0.04$~bpm as the standard deviation of $f[n]$. Spectrograms of raw signals at different levels of variation before being corrupted are shown in Fig.~\ref{fig::traceDynamic_example}. We observe a higher frequency energy diffusion when the trace variation increases, as the signal within each analysis window becomes less stationary.

We show the averaged system performance of the offline-AMTC in terms of \textsc{Rmse}, \textsc{ERate}, and \textsc{ECount} with respect to different combinations of the trace variation level and the selection of $k$ in the second row of Fig.~\ref{fig::impact_factor}. The SNR was fixed to $-10$~dB. From the plots, we observe that the performance decreases when the trace variation level gets higher, especially above level III. Even though the optimal selection of $k$ increases along with the trace variation level, \textsc{ERate} are controlled below $5\%$ when $k$ is fixed as $4$ or $6$ with trace variation level lower than V, suggesting the robustness of AMTC in terms of the trace variation level with a proper selection of the transitional probability parameter. The performance of the online-AMTC is slightly worse than that of the offline-AMTC as shown for an example of $k=4$ in Fig.~\ref{subfig::impact_factor_onoff_transitionMismatch}. The other observations about the offline-AMTC are also applicable to the online-AMTC.

\subsection{Impact of Trace Separation}\label{ssec::impact_trace_dis}
\begin{figure}
\centering
\subfigure[]{\includegraphics[width=0.7in]{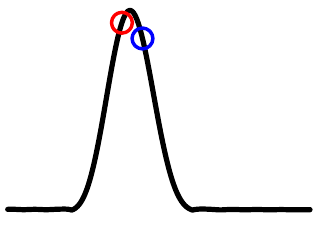}}%
\subfigure[]{\includegraphics[width=0.7in]{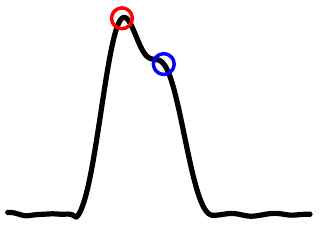}}%
\subfigure[]{\includegraphics[width=0.7in]{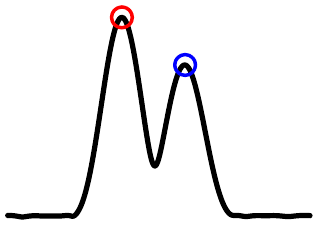}}%
\subfigure[]{\includegraphics[width=0.7in]{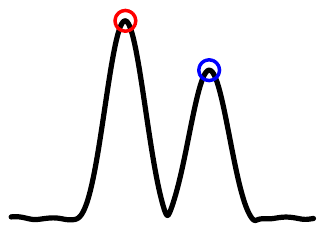}}%
\subfigure[]{\includegraphics[width=0.7in]{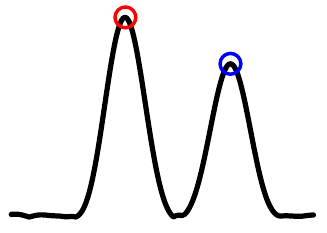}}%
\caption{Examples of the spectral frame when the Trace Relative Separation (TRS) equals $0.2$, $0.4$, $0.6$, $0.8$, and $1$, respectively from (a) to (e). Only part of the frame is displayed for better visualization.}
\label{fig::trace_dis_example}
\end{figure}

It is challenging for any frequency tracker to accurately distinguish and track two frequency traces that run very closely to each other, especially under low SNR conditions. To quantify the separation between two frequency components in a meaningful manner, we first defined a metric called Trace Relative Separation (TRS) as the ratio of the distance of two frequency components in the frequency domain to the mean width of their effective peaks. In Fig.~\ref{fig::trace_dis_example}, we show examples of the spectral distribution when $\text{TRS} = 0.2$, $0.4$, $0.8$, and $1$, respectively. 

We generated $200$ trials for each level of TRS using the same generative signal model described in Section~\ref{sssec::multitraceExp}. No unvoiced segment was added to the test signal and the TRS of two frequency traces was identical over time within each test signal. We show the averaged system performance of the offline-AMTC with respect to different levels of TRS and SNR in the last row of Fig.~\ref{fig::impact_factor}. From the plots, we know that the offline-AMTC is capable of tracking the frequency traces with \textsc{ERate} lower than $3\%$ when $\text{SNR} \leq-8$~dB, and $\text{TRS}\geq 0.4$. The estimation result when $\text{TRS} = 0.2$ is highly deviated from the ground truth. At this closeness level of TRS, more information or prior knowledge about the frequency components is expected to be incorporated to improve the estimation. The performance of the online-AMTC is slightly worse than that of the offline-AMTC with a similar trend in TRS to that of the offline-AMTC, an example of which is shown in Fig.~\ref{subfig::impatc_factor_onoff_closeTrace} when the SNR$=-8$~dB.

\section{Discussions}\label{section_discussion}
\subsection{Estimation of the Number of Traces}

In previous sections, we presented both the offline- and the online-AMTC algorithms with the assumption that the number of traces $L$ is known. In some cases, $L$ is unknown and needs to be estimated. Note that the process of estimating $L$ in the proposed AMTC system is equivalent to determining the number of iterations AMTC needs to take. The problem is then converted to deciding at which iteration should the AMTC stop. This problem can be solved by testing the hypothesis of the trace presence in the compensated spectrogram image $\mathbf{Z}_{(l)}$ at each iteration $l$.

\begin{figure}
\centering
\includegraphics[width=3.5in]{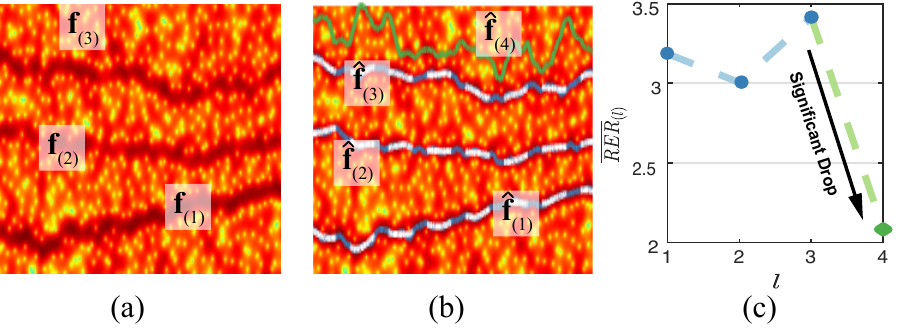}%
\caption{(a) Spectrogram image of a synthetic $-8$~dB signal with three frequency components and (b) the same image overlaid with ground-truth frequency components (white dashed line), the corresponding frequency estimates $\hat{\mathbf{f}}_{(1:3)}$(blue line) and one additional trace estimate $\hat{\mathbf{f}}_{(4)}$ (green line) using AMTC. (c) The corresponding averaged relative energy ratio $\overline{\text{RER}}$.}\label{fig_EstTraceNum}
\end{figure}

\begin{figure}
    \centering
    \subfigure[SNR$=-8$~dB]{\includegraphics[width=1.1in]{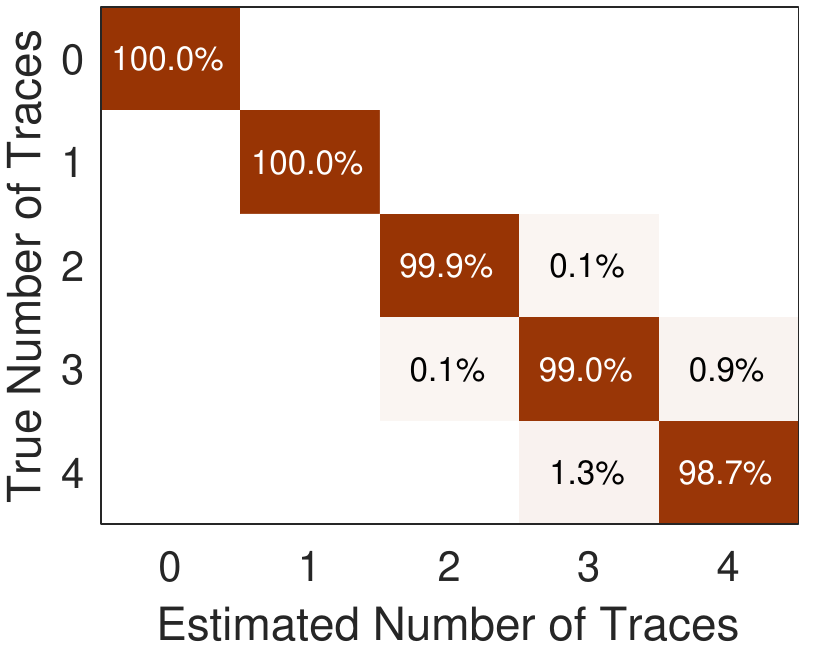}}
    \subfigure[SNR$=-10$~dB]{\includegraphics[width=1.1in]{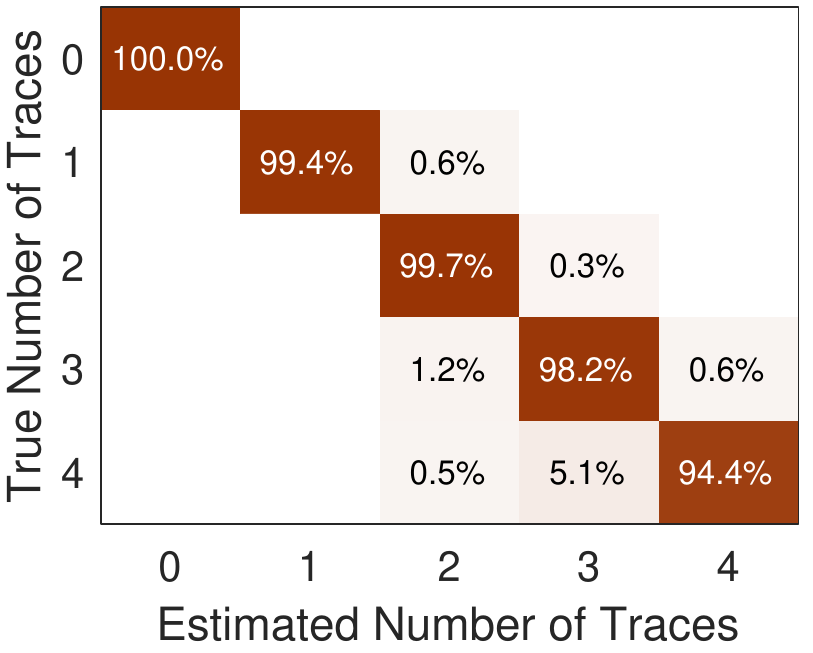}}
    \subfigure[SNR$=-12$~dB]{\includegraphics[width=1.1in]{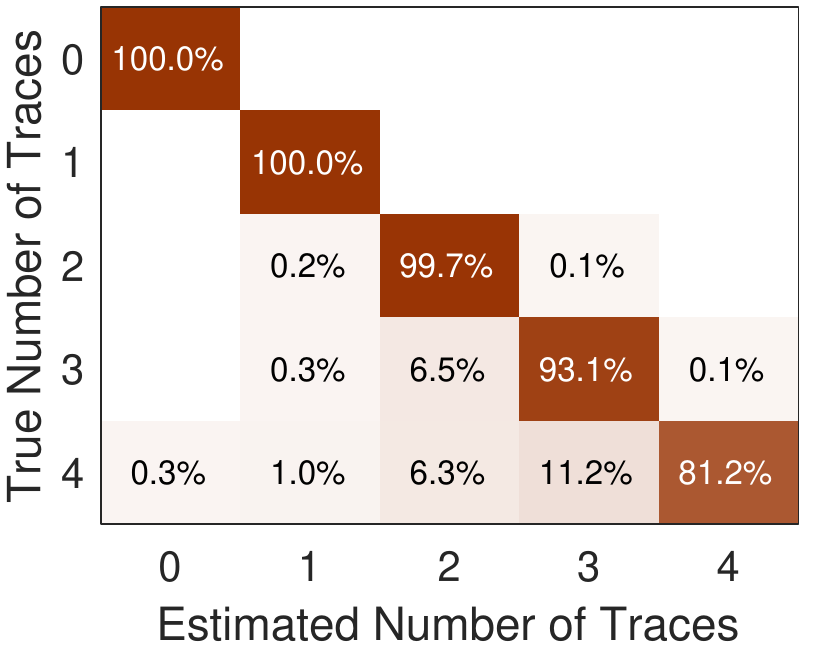}}
    \caption{Confusion matrix of the trace number estimator on different levels of SNR.}
    \label{fig:number_trace_confusion_mtx}
\end{figure}

In Section~\ref{subsection_TraceExistenceDetection}, we propose to use the RER measure to detect the presence of a frequency component in each frame. We are motivated by the fact that a low RER measure of a certain frame suggests low probability of the presence of a trace in that frame. Similarly, to test globally the trace presence at $l$th iteration of AMTC, we propose to evaluate the average of the statistics $\text{RER}_{(l)}$, namely, $\overline{\text{RER}}_{(l)}=\frac{1}{N}\sum_{n=1}^N \text{RER}_{(l)}(n)$. As one example shown in Fig.~\ref{fig_EstTraceNum}, the ground-truth number of traces in the spectrogram image is $3$. We observe a significant drop in $\overline{\text{RER}}_{(l)}$ from $l=3$ to $l=4$ in Fig.~\ref{fig_EstTraceNum}(c), when we run the offline-AMTC with four iterations. This observation coincides with the actual absence of the fourth trace. We therefore propose to estimate $L$ as $l-1$ if at the $l$th iteration, $\overline{\text{RER}}_{(l)}$ is below a preset threshold. The selection of the threshold is similar to the selection of $\Delta_{\text{RER}}$ discussed in Section~\ref{sssec::trace_detection}. To test the effectiveness of the propose detection method, we synthesized $2000$ signals using the generative model introduced in Section~\ref{sssec::multitraceExp} with equal numbers of signals which contained zero, one, two, three, and four frequency traces. The detection accuracy under three levels of SNR is shown in Fig.~\ref{fig:number_trace_confusion_mtx} in the form of confusion matrices. We observe highly accurate results with $99.7\%$ detection accuracy when the SNR equals $-12$~dB and the number of traces is no more than three in this experiment setting.

\subsection{Signals with Multiple Harmonics}
When multiple harmonic traces appear in the spectrogram (\textit{e.g.}, audio signals, Electrocardiography (ECG) signals), the AMTC algorithms may extract several harmonic traces that originated from one single source. Take the human speech signal as an example. The fundamental frequency range of interest, 85 Hz to 255 Hz \cite{titze1998principles, baken2000clinical}, may cover both fundamental frequency components as well as second-order harmonics. For example, a peak in 200 Hz can be considered as the fundamental frequency component of a female speaker, or it can also represent the second-order harmonic of a male speaker. In this regard, the STFT spectrum feature might not be considered as a proper input of a robust fundamental frequency tracker. Instead, this problem can be addressed by introducing several alternative robust spectral features, \textit{e.g.}, the subharmonic summation method \cite{hermes1988measurement}, the discrete logarithmic Fourier transform \cite{wang2000robust}, and the frequency autocorrelation function \cite{zahorian2008spectral}. Similar to the idea of harmonic combining algorithm \cite{hajj2013spectrum} used for ENF case, these methods are capable of combining harmonic spectral features and improving the SNR of the fundamental frequency. The tracking performance is therefore expected to be better by feeding in any of these three features rather than the STFT spectrogram.

\begin{figure}
\centering
\includegraphics[width=3.5in]{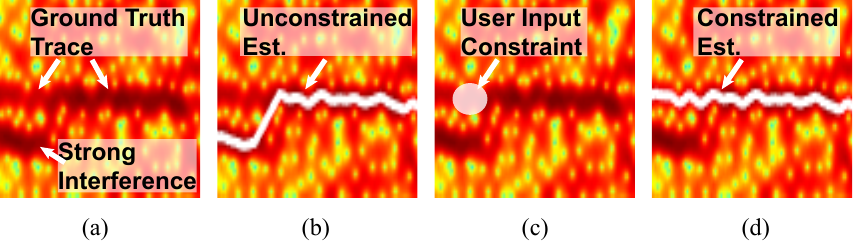}%
\caption{(a) Spectrogram of a synthetic signal with ground-truth frequency around $95$~bpm and strong nearby interference from $0$--$0.4$~min. (b) Unconstrained trace estimate. (c) Spectrogram overlaid with user input constraint (in semi-transparent white circle).  (d) Constrained trace estimate. }\label{fig_UserInput}
\end{figure}

\subsection{Accommodating Human-in-the-Loop Interactions}
AMTC has its limitations in some specific cases. Due to the greedy nature of the searching strategy in each iteration, the algorithm may find incorrect traces when nearby strong interference is present, or two traces with similar energies runs closely in time. We show in Fig.~\ref{fig_UserInput}(b) an example that strong interference near the ground-truth frequency trace can make it challenging for AMTC to find the correct trace. Without extra information, even a human observer can make mistakes in this scenario.
For some applications when the analysis is performed offline and people have some prior knowledge about the trace shape or the trace frequency range, it is beneficial to allow users to input high-level cues~\cite{avidan2007seam,chu2013halftone} to guide our proposed estimator's priority to find the correct trace. As an example, Fig.~\ref{fig_UserInput}(c) shows a constraint provided by a user using a semi-transparent white circle for an estimated trace to pass through. Fig.~\ref{fig_UserInput}(d) shows the constrained estimation result, which is achieved by scaling up the spectrum entries in the constraint region until the estimated trace passed through the region. The constrained tracking result reveals that AMTC correctly captured the true trace by shifting its attention from interference to the user-defined region.

\section{Conclusions}\label{section_conclude}
In this paper, we have addressed the problem of tracking multiple weak frequency components from a time-frequency representation of the system's preprocessing results (such as a spectrogram), and proposed both offline and online versions of a new trace detection and tracking algorithm called AMTC. By iteratively and adaptively estimating dynamic traces through forward and backward passes, AMTC can provide accurate estimates even for weak frequency traces. Extensive experiments using both synthesis and real-world data reveal that the proposed method outperforms several representative prior methods under low SNR conditions and can be implemented in near real-time settings. The effectiveness of the proposed algorithm can empower the development of new frequency-based forensic technologies and other small-signal applications.

\section*{Acknowledgment}
The authors would like to thank the associate editor and anonymous reviewers whose comments and suggestions have helped improve the clarity and evaluations of this paper.
\ifCLASSOPTIONcaptionsoff
  \newpage
\fi

\bibliographystyle{IEEEtran}
\bibliography{AMTC_ref}

\begin{IEEEbiography}[{\includegraphics[width=1in,height=1.25in,clip,keepaspectratio]{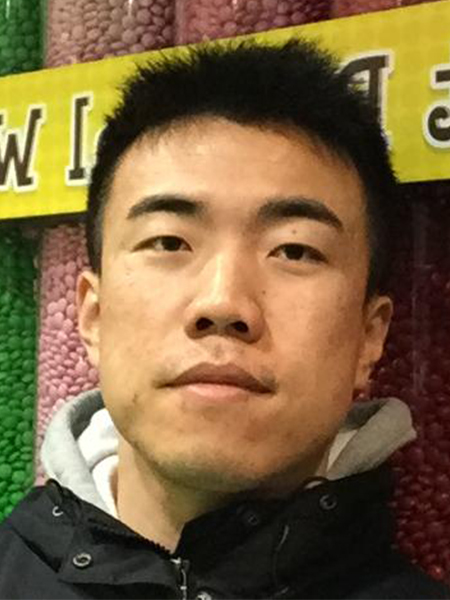}}]{Qiang Zhu} (S'17) received his B.E. degree from Zhejiang University, Hangzhou, China, in 2010, his M.S. degree in control science and engineering from Shanghai Jiao Tong Unversity, Shanghai, China, in 2014, and his Ph.D. degree in electrical engineering from the University of Maryland, College Park, USA, in 2020. He is a research scientist at Facebook since 2020. His research interests are signal processing, machine learning, and information retrieval. He received the Distinguished Teaching Assistance Award in 2016 from the University of Maryland.
\end{IEEEbiography}

\begin{IEEEbiography}[{\includegraphics[width=1in,height=1.25in,clip,keepaspectratio]{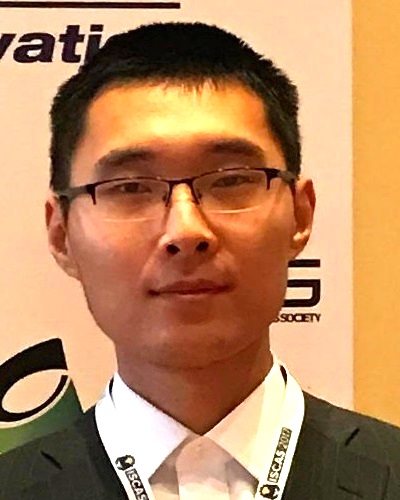}}]{Mingliang Chen} (S'18) received his B.E. and M.S. degree in electronic information engineering from Shanghai Jiao Tong University, China, in 2013 and 2016, respectively. He is currently pursuing the Ph.D. degree in electrical and computer engineering at the University of Maryland, College Park. His current research interests are signal and image processing and machine learning. He received the 2019 Jimmy H.C. Lin Award for Innovation from the University of Maryland.
\end{IEEEbiography}

\begin{IEEEbiography}[{\includegraphics[width=1in,height=1.25in,clip,keepaspectratio]{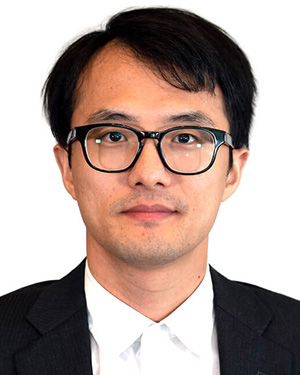}}]{Chau-Wai Wong}
(S'05--M'16) received his B.Eng. and M.Phil. degrees in electronic and information engineering from The Hong Kong Polytechnic University in 2008 and 2010, and the Ph.D. degree in electrical engineering from the University of Maryland, College Park in 2017. He is currently an Assistant Professor at the Department of Electrical and Computer Engineering and the Forensic Sciences Cluster, North Carolina State University. His research interests include multimedia forensics, statistical signal processing, machine learning, data analytics, and video coding. Dr. Wong received a Top-4 Student Paper Award, Future Faculty Fellowship, HSBC Scholarship, and Hitachi Scholarship. He was involved in organizing the third edition of the IEEE Signal Processing Cup in 2016 on electric network frequency forensics.
\end{IEEEbiography}

\begin{IEEEbiography}[{\includegraphics[width=1in,height=1.25in,clip,keepaspectratio]{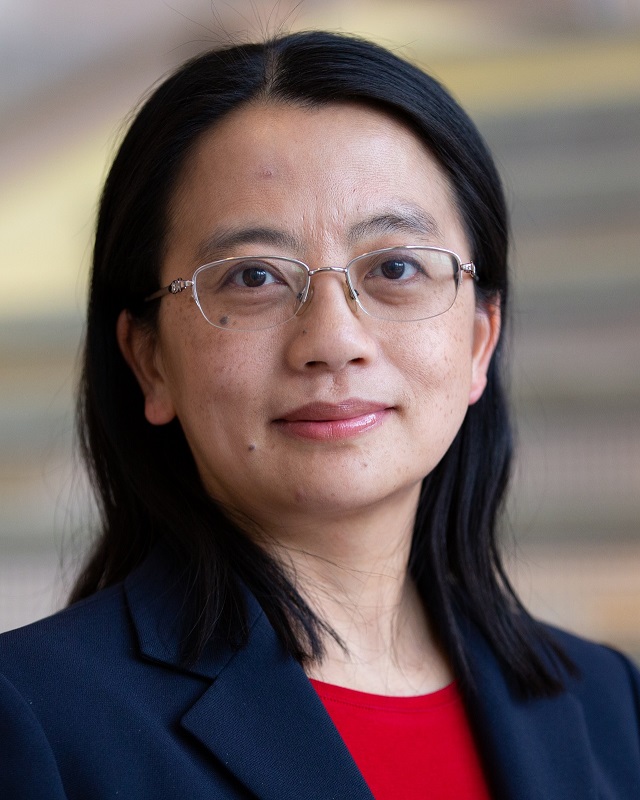}}]{Min Wu} (S'95--M'01--SM'06--F'11) received her Ph.D. degree in electrical engineering from Princeton University. Since 2001, she has been with the University of Maryland at College Park, where she is currently a Professor and Associate Dean of Engineering, and a University Distinguished Scholar-Teacher. Her research interests include information security and forensics, multimedia signal processing, and applications of data science and machine learning in health and IoT. She is a Fellow of the IEEE, AAAS, and the U.S. National Academy of Inventors.
\end{IEEEbiography}

\end{document}